\documentclass[12pt]{article}
\usepackage[margin=1in]{geometry}
\usepackage[utf8]{inputenc}

\usepackage{amsmath,amssymb,amsfonts,amsthm}
\usepackage{graphicx}
\usepackage{xcolor}
\usepackage{cite}
\usepackage{url}
\usepackage{hyperref}
\hypersetup{colorlinks=false,linkcolor=black}
\usepackage[numbers,sort&compress]{natbib}

\allowdisplaybreaks
\usepackage{tabularx}
\usepackage{bbm}
\usepackage{listings}
\usepackage{longtable}
\newcommand\refeq[1]{Eq.~(\ref{#1})}

\newcommand\refta[1]{Tab.~\ref{#1}}
\newcommand\refse[1]{Sect.~\ref{#1}}

\newcommand\citere[1]{Ref.~\cite{#1}}
\newcommand\citeres[1]{Refs.~\cite{#1}}

\def\reffi#1{\mbox{Fig.~\ref{#1}}}
\newcommand\refli[1]{Listing~(\ref{#1})}

\newcommand{\vil}{\widetilde{\nu}_{iL}^\mathcal{R}}

\newcommand{\at}{A_t}
\newcommand{\yt}{Y_t}

\newcommand{\yv}{Y^\nu_{ii}}
\newcommand{\av}{A^\nu_{ii}}

\newcommand{\mst}{m_{\widetilde{t}_1}^2}
\newcommand{\mstt}{m_{\widetilde{t}_2}^2}

\newcommand{\gev}{\ \mathrm{GeV}}
\newcommand{\tev}{\ \mathrm{TeV}}

\newcommand\DRbar{\ensuremath{\smash{\overline{\mathrm{DR}}}}}
\newcommand\MSbar{\ensuremath{\smash{\overline{\mathrm{MS}}}}}
\newcommand{\ii}{\text{i}}

\begin{document}

\thispagestyle{empty}

\def\thefootnote{\fnsymbol{footnote}}

\begin{flushright}
  DESY-20-162 ~~
\end{flushright}

\vspace*{2cm}

\begin{center}
  {\Large \textbf{\texttt{munuSSM}: A python package for the $\mu$-from-$\nu$
    Supersymmetric Standard Model} }  
  
  \vspace{1cm}

  Thomas Biekötter\footnote{thomas.biekoetter@desy.de}\\
  {\textit{  
      DESY,  Notkestra{\ss}e  85,  22607  Hamburg,  Germany\\[0pt]
  }}
  
\vspace*{1cm}

\begin{abstract}
We present the public python package \texttt{munuSSM} that can be
used for phenomenological studies in the context of
the $\mu$-from-$\nu$ Supersymmetric Standard Model
($\mu\nu$SSM). The code incorporates the radiative corrections
to the neutral scalar potential at full one-loop level.
Sizable higher-order corrections, required for an accurate prediction
of the SM-like Higgs-boson mass, can be consistently included via an automated
link to the public code \texttt{FeynHiggs}.
In addition, a calculation of effective couplings and branching ratios
of the neutral and charged Higgs bosons is implemented.
This provides the required
ingredients to check a benchmark point against collider constraints
from searches for additional Higgs bosons via an interface
to the public code \texttt{HiggsBounds}. 
At the same time, the signal rates of the SM-like Higgs boson
can be tested applying the experimental results implemented
in the public code \texttt{HiggsSignals}.
The python package is constructed in a flexible and modular way, such
that it provides a simple framework that can be extended by
the user with further calculations of
observables and constraints on the model parameters.
\end{abstract}

\end{center}

\vspace*{1cm}

The source code of \texttt{munuSSM}
and instructions for the installation are
available at:
\begin{center}
\url{https://gitlab.com/thomas.biekoetter/munussm}
\end{center}
\newpage

\renewcommand{\thefootnote}{\arabic{footnote}}
\setcounter{footnote}{0} 

\section{Introduction}
\label{intro}
Supersymmetry (\textsc{Susy}) is one of the prime
candidates for physics Beyond the Standard Model (BSM).
A particularly well motivated \textsc{Susy} model
is the so-called $\mu$-from-$\nu$ Supersymmetric
Standard Model
($\mu\nu$SSM)~\cite{Bratchikov:2005vp,Munoz:2009an}.
Beyond the usual benefits
of low-scale \textsc{Susy}, i.e., providing a
solution to the hierarchy problem and allowing for
the unification of the three Standard Model (SM)
gauge couplings, the $\mu\nu$SSM incorporates an
electroweak seesaw mechanism. Without introducing
any scales beyond the \textsc{Susy}-breaking scale
$M_S$, the tiny neutrino masses and their mixing
pattern can be accommodated assuming neutrino
Yukawa couplings $Y^\nu$ of the order of the
electron Yukawa coupling.
Furthermore, the superpotential is
scale invariant and the $\mu$-term of the
superpotential of the Minimal Supersymmetric
Standard Model (MSSM) is generated dynamically
during Electroweak Symmetry Breaking (EWSB).
Apart from the Higgs doublets, also
the scalar partners of the
neutrinos (called sneutrinos)
obtain a Vacuum Expectation Value (vev)
during EWSB.
The tree-level mass of the SM-like Higgs boson
receives additional contributions stemming from
portal couplings between the Higgs doublet fields
and the right-handed sneutrinos. Thus,
compared to the MSSM, a value of $\sim 125\gev$
can be achieved with fewer radiative
corrections.

The $\mu\nu$SSM is especially interesting in view
of EWSB. In this model the stability
of the proton is assured by forbidding operators
breaking baryon number~\cite{Munoz:2009an}.
However, it does not
assume $R$-parity conservation and the breaking of
lepton number is induced via terms
proportional to $Y^\nu$ by construction.
Thus, the left- and right-handed sneutrinos mix
with the neutral scalar components of the
Higgs doublet superfields.
Apart from that, the scalar partners of the leptons
(called sleptons)
mix with the charged scalar components of the
Higgs doublet superfields.
Neglecting CP violation, as
we will do throughout this paper, the Higgs sector
of the $\mu\nu$SSM consists of a total of
8 CP-even, 7 CP-odd, and  ${2\times 7}$ charged
scalars. In addition, there are the usual
pseudoscalar and charged Goldstone bosons
$G^0$ and $G^\pm$.
During EWSB all of the 8 neutral scalar fields
obtain a vev. While the mixing of the doublet Higgs
bosons and the gauge-singlet right-handed sneutrinos
can in principle be arbitrarily large, the mixing
between the left-handed sneutrinos and the Higgs
doublets is suppressed by the small values
of $Y^\nu$. This decoupling is also reflected in
a large hierarchy between the vevs. The
vevs of the Higgs doublets $v_u$ and $v_d$ and
the vevs of the right-handed sneutrinos $v_{iR}$ ($i=1,2,3$)
are related to the breaking scale of the EW
symmetry and \textsc{Susy}. The vevs of the
left-handed sneutrinos $v_{iL}$, on the other
hand, are related to
the breaking of lepton number, and therefore
suppressed by a factor of $Y^\nu$ compared to
$v_d$, $v_u$ and $v_{iR}$.

These unique features motivated the precise analysis
of the Higgs sector of the model, including
the radiative corrections at full
one-loop level. At first, we studied a simplified version
of the $\mu\nu$SSM with a single right-handed
neutrino superfield~\cite{Biekotter:2017xmf}. Later on,
we extended the calculation to the complete model
with three right-handed neutrino
superfields~\cite{Biekotter:2019gtq}.
In the latter, three non-zero
left-handed neutrino masses can already be
accommodated at tree level.
It was found that for the SM-like Higgs-boson
mass the mixing effects between doublet fields
and right-handed sneutrinos are important
at loop level and have to be taken into
account, while the tiny mixing with the left-handed
sneutrinos does not play a role.
However, the left-handed sneutrinos themselves
are subject to potentially large corrections
proportional to $Y^\nu A_t Y_t / v_{iL}$, in which
the suppression of the factors $Y^\nu$ is compensated
by the left-handed vevs in the denominator and
a factor of the soft scalar top (called stop) mixing parameter~$A_t$
times the top Yukawa coupling~$Y_t$.

It is known from the MSSM that corrections
to the Higgs-boson mass beyond one-loop level
are sizable and have to be taken into
account~\cite{Zhang:1998bm,Espinosa:1999zm,Heinemeyer:1998np}.
These higher-order contributions should be
included in an approximate form also in the
$\mu\nu$SSM in order to obtain an accurate
prediction.
Combining the higher-order effects with the
full one-loop result,
it was shown that the $\mu\nu$SSM can
easily accommodate a Higgs boson at $\sim 125\gev$
that reproduces the measured signal rates
within the current experimental
uncertainties~\cite{Biekotter:2019gtq}.
Apart from that, interesting new Higgs physics can
be realized at relatively low masses, since
the right- and the left-handed sneutrinos
could have escaped discovery so far even for
masses below $125\gev$~\cite{Biekotter:2019gtq}.
Note that the right-handed
sneutrinos are gauge singlets, such that they
naturally have reduced couplings to the SM
particles. In fact, they only couple to the
SM via the mixing with the doublet fields, for instance
the SM-like Higgs boson. Such a scenario is particularly interesting,
as it can be probed not only by directly
searching for additional Higgs bosons, but also
indirectly by measuring possible deviations
from the SM predictions
of the couplings of the Higgs boson at
$125\gev$~\cite{Biekotter:2019gtq,Kpatcha:2019qsz}.
A possible detection of light left-handed sneutrinos
requires dedicated searches when they are the
lightest \textsc{Susy} particle, since their
decay must proceed via $R$-parity violating
couplings~\cite{Ghosh:2017yeh,Lara:2018rwv,Kpatcha:2019gmq}.

In this paper we present the tool
\texttt{munuSSM} for the phenomenological
study of the $\mu\nu$SSM.
In contrast to the already existing public codes
\texttt{SARAH/SPheno}~\cite{Porod:2003um,Porod:2011nf,Staub:2009bi,
Staub:2010jh,Staub:2012pb,Staub:2013tta} and
\texttt{FlexibleSUSY}~\cite{Athron:2014yba,Athron:2016fuq,Athron:2017fvs},
which are designed to be generically applicable to
various different (\textsc{Susy}) extensions of the SM,
the code \texttt{munuSSM}
is targeted specifically at the $\mu\nu$SSM.
First and foremost, it makes
the one-loop corrections to the Higgs potential
publicly available in terms of the momentum-dependent
renormalized scalar self energies. These are used in
combination with leading higher-order corrections
from the public code
\texttt{FeynHiggs}~\cite{Heinemeyer:1998yj,Heinemeyer:1998np,
Degrassi:2002fi,Frank:2006yh,Hahn:2013ria,Bahl:2016brp,
Bahl:2017aev,Bahl:2018qog}
to accurately predict the particle masses of the
neutral scalars, in particular the SM-like Higgs-boson
mass. In addition, the radiative corrections to the
mixing matrix elements are used to obtain effective
couplings of the scalars to the SM particles.
Furthermore, the calculation of the branching ratios
of the neutral and charged scalars is
implemented. For decays into SM particles, the branching
ratios are obtained by rescaling the SM
predictions~\cite{Heinemeyer:2013tqa,deFlorian:2016spz} by the
effective couplings. For decays to BSM final states, the
branching ratios are calculated from scratch at leading order,
however including radiative corrections by rotating the
tree-level couplings into the loop-corrected mass eigenstate basis.
The effective couplings and branching ratios can be
directly interfaced to the public codes
\texttt{HiggsBounds}~\cite{Bechtle:2008jh,Bechtle:2011sb,
Bechtle:2013gu,Bechtle:2013wla,Bechtle:2015pma,Bechtle:2020pkv}
and \texttt{HiggsSignals}~\cite{Bechtle:2013xfa,
Stal:2013hwa,Bechtle:2014ewa,Bechtle:2020uwn}
to test a benchmark point against collider constraints.
The interface to \texttt{HiggsBounds} also provides
the LHC cross sections normalized to the SM prediction,
which are extracted from the effective couplings.

The paper is organized as follows. We start by
briefly introducing the model in \refse{intromunu}.
The overall structure of the code \texttt{munuSSM}
and its subpackages
are described in \refse{secmunu}, paying special attention
to the links to other public codes in \refse{dringesichert}.
Afterwards, we explain the installation process and the
basic user instructions in \refse{insta} and \refse{usage}.
A simple example analysis is described in \refse{secexample}.
We conclude in \refse{conclu}.

\section{The $\mu$-from-$\nu$ Supersymmetric Standard Model}
\label{intromunu}
In this section we provide the basic definitions and
conventions under which the model predictions were
implemented. A more detailed motivation and a review
of the $\mu\nu$SSM can be found in \citere{Lopez-Fogliani:2020gzo}.
The calculation of the radiative corrections
to the Higgs potential are described in detail in
\citeres{Biekotter:2017xmf,Biekotter:2019gtq}.

The superpotential of the $\mu\nu$SSM is written as
\begin{align}
W = \; & 
\epsilon_{ab} \left(
Y^e_{ij} \, \hat H_d^a\, \hat L^b_i \, \hat e_j^c +
Y^d_{ij} \, \hat H_d^a\, \hat Q^{b}_{i} \, \hat d_{j}^{c} 
+
Y^u_{ij} \, \hat H_u^b\, \hat Q^{a}_{i} \, \hat u_{j}^{c}
\right)
\nonumber\\
+ \; &   
\epsilon_{ab} \left(
Y^{\nu}_{ij} \, \hat H_u^b\, \hat L^a_i \, \hat \nu^c_j 
-
\lambda_{i} \, \hat \nu^c_i\, \hat H_u^b \hat H_d^a
\right)
+
\frac{1}{3}
\kappa{_{ijk}} 
\hat \nu^c_i\hat \nu^c_j\hat \nu^c_k
\ ,
\label{superpotential}
\end{align}
where $\hat H_d^T=(\hat H_d^0, \hat H_d^-)$ and 
$\hat H_u^T=(\hat H_u^+, \hat H_u^0)$ are the Higgs doublet
superfields, $\hat Q_i^T=(\hat u_i, \hat d_i)$ and 
$\hat L_i^T=(\hat \nu_i, \hat e_i)$ are the left-chiral
quark and lepton superfields,
and $\hat u_{j}^{c}$, $\hat d_{j}^{c}$, $\hat e_j^c$ and $\hat{\nu}^c$ are
the right-chiral quark and lepton superfields.
$i,j=1,2,3$ are the family indices, and 
$a,b=1,2$ are indices of the fundamental representation of SU(2) 
with $\epsilon_{ab}=1$.
The colour indices are not written out.

In the framework of low-energy \textsc{Susy} breaking,
the soft Lagrangian of the $\mu\nu$SSM is given
by~\cite{Brignole:1997dp}
\begin{align}
-\mathcal{L}_{\text{soft}}  & = \;  
\epsilon_{ab} \left(
T^e_{ij} \, H_d^a  \, \widetilde L^b_{iL}  \, \widetilde e_{jR}^* +
T^d_{ij} \, H_d^a\,   \widetilde Q^b_{iL} \, \widetilde d_{jR}^{*} 
+
T^u_{ij} \,  H_u^b \widetilde Q^a_{iL} \widetilde u_{jR}^*
+ \text{h.c.}
\right)
\nonumber \\
+ \; &
\epsilon_{ab} \left(
T^{\nu}_{ij} \, H_u^b \, \widetilde L^a_{iL} \widetilde \nu_{jR}^*
- 
T^{\lambda}_{i} \, \widetilde \nu_{iR}^*
\, H_d^a  H_u^b
+ \frac{1}{3} T^{\kappa}_{ijk} \, \widetilde \nu_{iR}^*
\widetilde \nu_{jR}^*
\widetilde \nu_{kR}^*
\
+ \text{h.c.}\right)
\nonumber\\
+ \; & 
\left(m_{\widetilde{Q}}^2\right)_{ij} 
\widetilde{Q}_{iL}^{a*}
\widetilde{Q}^a_{jL}
+\left(m_{\widetilde{u}}^{2}\right)_{ij} \widetilde{u}_{iR}^*
\widetilde u_{jR}
+ \left(m_{\widetilde{d}}^2\right)_{ij}  \widetilde{d}_{iR}^*
\widetilde d_{jR}
+
\left(m_{\widetilde{L}}^2\right)_{ij}  
\widetilde{L}_{iL}^{a*}  
\widetilde{L}^a_{jL}
\nonumber\\
+ \; &
\left(m_{\widetilde{\nu}}^2\right)_{ij} \widetilde{\nu}_{iR}^*
\widetilde\nu_{jR} 
+
\left(m_{\widetilde{e}}^2\right)_{ij}  \widetilde{e}_{iR}^*
\widetilde e_{jR}
+ 
m_{H_d}^2 {H^a_d}^*
H^a_d + m_{H_u}^2 {H^a_u}^*
H^a_u \\
+ \; & \left(
\left( m_{H_d\widetilde{L}}^2 \right)_{i}H_d^{a*} \widetilde{L}_{iL}^a
+ \text{h.c.} \right) \nonumber \\
+ \; &  \frac{1}{2}\, \left(M_3\, {\widetilde g}\, {\widetilde g}
+
M_2\, {\widetilde{W}}\, {\widetilde{W}}
+M_1\, {\widetilde B}^0 \, {\widetilde B}^0 + \text{h.c.} \right)\ .
\label{Vsoft}
\end{align}
The parameters $m_{H_d\widetilde{L}}^2$ are absent at tree level,
as they are non-diagonal in field space. However, in the code
the terms are taken into account, because
they are required for the renormalization of the Higgs potential
(see \citeres{Biekotter:2017xmf,Biekotter:2019gtq} for details).
In addition, flavour mixing is neglected in the quark and the
squark sector, such that the corresponding soft mass parameters
only have diagonal non-zero entries 
$m_{\widetilde{Q}_{i} }^2$, $m_{\widetilde{u}_{i}}^2$ and
$m_{\widetilde{d}_{i}}^2$. The soft trilinear couplings
are written as $T^u_i=A^u_i Y^u_i$, $T^d_i=A^d_i Y^d_i$,
where $Y^u_i$ and $Y^d_i$ are the diagonal entries of
the Yukawa couplings of the up- and down-type quarks
and no summation over repeated indices is implied.
In the lepton sector, the flavour symmetries are broken
automatically after EWSB. Thus, we decompose the
soft trilinear couplings as
$T^e_{ij}=A^e_{ij}Y^e_{ij}$ and
$T^\nu_{ij}=A^\nu_{ij} Y^\nu_{ij}$,
again without summation over repeated indices.
The lepton-flavour mixing is suppressed
by factors of $Y^\nu_{ij}$ and therefore only
sizable for the light left-handed neutrinos.
Finally, we write the portal coupling and the
self coupling of the right-handed sneutrinos
as $T^\lambda_i = A^\lambda_i \lambda_i$
and $T^\kappa_{ijk} = A^\kappa_{ijk} \kappa_{ijk}$,
noting that both $\kappa_{ijk}$ and $A^\kappa_{ijk}$
are symmetric under the exchange of indices.

The soft terms together with the $D$-term and $F$-term
contributions from the superpotential define
the tree-level neutral scalar potential
\begin{equation}
V^{(0)} = V_{\text{soft}} + V_F  +  V_D\ , 
\label{finalpotential}
\end{equation}
with
\begin{align}
V_{\text{soft}}= \; & 
\left(
T^{\nu}_{ij} \, H_u^0\,  \widetilde \nu_{iL} \, \widetilde \nu_{jR}^* 
- T^{\lambda}_{i} \, \widetilde \nu_{iR}^*\, H_d^0  H_u^0
+ \frac{1}{3} T^{\kappa}_{ijk} \, \widetilde \nu_{iR}^* \widetilde \nu_{jR}^* 
\widetilde \nu_{kR}^*\
+
\text{h.c.} \right)
\nonumber\\
+ \; &
\left(m_{\widetilde{L}}^2\right)_{ij} \widetilde{\nu}_{iL}^* \widetilde\nu_{jL}
+
\left(m_{\widetilde{\nu}}^2\right)_{ij} \widetilde{\nu}_{iR}^* \widetilde\nu_{jR} +
m_{H_d}^2 {H^0_d}^* H^0_d + m_{H_u}^2 {H^0_u}^* H^0_u
\ ,
\label{akappa}
\\
\nonumber
\\
V_{F}  = \; &
 \lambda_{j}\lambda_{j} H^0_{d}H_d^0{^{^*}}H^0_{u}H_u^0{^{^*}}
 +
\lambda_{i}\lambda_{j} \tilde{\nu}^{*}_{iR}\tilde{\nu}_{jR}H^0_{d}H_d^0{^*}
 +
\lambda_{i}\lambda_{j}
\tilde{\nu}^{*}_{iR}\tilde{\nu}_{jR}  H^0_{u}H_u^0{^*}   
\nonumber\\                                              
+ \; &
\kappa_{ijk}\kappa_{ljm}\tilde{\nu}^*_{iR}\tilde{\nu}_{lR}
                                   \tilde{\nu}^*_{kR}\tilde{\nu}_{mR}
- \left(\kappa_{ijk}\lambda_{j}\tilde{\nu}^{*}_{iR}\tilde{\nu}^{*}_{kR} H_d^{0*}H_u^{0*}                                      
 -Y^{\nu}_{ij}\kappa_{ljk}\tilde{\nu}_{iL}\tilde{\nu}_{lR}\tilde{\nu}_{kR}H^0_{u}
 \right.
 \nonumber\\
 + \; &
 \left.
 Y^{\nu}_{ij}\lambda_{j}\tilde{\nu}_{iL} H_d^{0*}H_{u}^{0*}H^0_{u}
+{Y^{\nu}_{ij}}\lambda_{k} \tilde{\nu}_{iL}^{*}\tilde{\nu}_{jR}\tilde{\nu}_{kR}^* H^0_{d}
 + \text{h.c.}\right) 
\nonumber \\
+ \; &
Y^{\nu}_{ij}{Y^{\nu}_{ik}} \tilde{\nu}^{*}_{jR}
\tilde{\nu}_{kR}H^0_{u}H_u^0{^*}                                                
 +
Y^{\nu}_{ij}{Y^{\nu}_{lk}}\tilde{\nu}_{iL}\tilde{\nu}_{lL}^{*}\tilde{\nu}_{jR}^{*}
                                  \tilde{\nu}_{kR}  
 +
Y^{\nu}_{ji}{Y^{\nu}_{ki}}\tilde{\nu}_{jL}\tilde{\nu}_{kL}^* H^0_{u}H_u^{0*}\, ,
\\
\nonumber
\\
V_D  = \; &
\frac{1}{8}\left(g_1^{2}+g_2^{2}\right)\left(\widetilde\nu_{iL}\widetilde{\nu}_{iL}^* 
+H^0_d {H^0_d}^* - H^0_u {H^0_u}^* \right)^{2}\, .
\label{dterms}
\end{align}
During EWSB the neutral scalar
fields acquire a vev. We use the decomposition
\begin{align}
H_d^0 &= \frac{1}{\sqrt 2} \left(H_{d}^{\mathcal{R}} + v_d +
\ii\, H_{d}^{\mathcal{I}}\right)\ ,
\label{vevd} \\
H^0_u &= \frac{1}{\sqrt 2} \left(H_{u}^{\mathcal{R}}  + v_u +
  \ii\, H_{u}^{\mathcal{I}}\right)\ ,  
\label{vevu} \\
\widetilde{\nu}_{iR} &=
      \frac{1}{\sqrt 2} \left(\widetilde{\nu}^{\mathcal{R}}_{iR}+ v_{iR} +
        \ii\, \widetilde{\nu}^{\mathcal{I}}_{iR}\right) \ ,
\label{vevnuc} \\
\widetilde{\nu}_{iL} &= \frac{1}{\sqrt 2} \left(\widetilde{\nu}_{iL}^{\mathcal{R}} 
  + v_{iL} +
    \ii\, \widetilde{\nu}_{iL}^{\mathcal{I}}\right)\ ,
\label{vevnu}
\end{align}
such that the vevs are given by\footnote{We will refer to
the parameters $v_u$, $v_d$, $v_{iL}$ and $v_{iR}$ as
vevs interchangeably.}
\begin{align}
\langle H_d^0 \rangle = \frac{v_d}{\sqrt 2}\ , \, \quad 
\langle H_u^0 \rangle = \frac{v_u}{\sqrt 2}\ , \,
\quad 
\langle \widetilde \nu_{iR}\rangle = \frac{v_{iR}}{\sqrt 2}\ , \,  \quad
\langle \widetilde \nu_{iL} \rangle = \frac{v_{iL}}{\sqrt 2}
\; .
\label{vevs}
\end{align}
The subscripts $^{\mathcal{R}}$ and
$^{\mathcal{I}}$ denote CP-even and -odd
components of each scalar field,
respectively. To make a connection to the SM and
the MSSM, we define the parameters
\begin{equation}
v^2 = v_u^2 + v_d^2 + v_{iL} v_{iL} \sim 246\gev \quad
\text{and} \quad \tan\beta = \frac{v_u}{v_d} \ .
\end{equation}
As already mentioned in \refse{intro},
the size of the left-handed vevs $v_{iL}$ is
suppressed by factors of $Y^\nu_{ij}$ compared
to the other vevs. Hence, they are of the order
of $\sim 10^{-5}$ to $10^{-4}\gev$.
The minimization or tadpole equations relate
the soft mass parameters to the vevs.
For numerical reasons it is most convenient to use the vevs as input
parameters and solve the tadpole equations for
the soft masses squared $m_{H_d}^2$, $m_{H_u}^2$,
$(m_{\widetilde{L}}^2)_{ii}$ and
$(m_{\widetilde{\nu}}^2)_{ii}$.
The precise form of the tadpole equations can
be found in \citere{Biekotter:2019gtq}.

The expressions for the tree-level masses of all particles of the
model in terms of the parameters defined above can
be found in \citere{Biekotter:2019gtp}.
Since they are rather lengthy
we do not repeat them here. The expressions for
the tree-level couplings
of the particles are even larger due to the complicated
mixing in the scalar sector, such that we do not
state them either. Instead, we provide a
\texttt{FeynArts}~\cite{Hahn:2000kx} modelfile upon request
that contains the couplings in the 't Hooft-Feynman gauge
in \texttt{Mathematica} syntax.\footnote{The couplings of the gravitino and
the axino, both potential dark matter candidates in
the $\mu\nu$SSM~\cite{Gomez-Vargas:2019vci,Gomez-Vargas:2019mqk},
are not included. However, they only
play a role for the DM phenomenology and are irrelevant
for the Higgs and collider physics.}
The modelfile was initially created with the public tool
\texttt{SARAH}~\cite{Staub:2013tta}, but further modified by hand to
allow the usage of the tool
\texttt{FormCalc}~\cite{Hahn:1998yk}, which
by default cannot process the huge expressions for the
couplings produced by \texttt{SARAH}.

In the code \texttt{munuSSM}, the calculation of the
tree-level spectrum and the corresponding mixing matrices,
as well as the tree-level couplings, are evaluated in
Fortran subroutines. This allows for a larger floating-point
precision, which is necessary
due to the tiny $R$-parity violating
mixing effects and the large hierarchy between the masses
in the neutral fermion sector.
Apart from that, the usage of Fortran vastly improves
the running time compared to an implementation in python.
We note that the running time is currently dominated by the
calculation of the complete set of tree-level couplings.

\subsection{Radiative corrections in the Higgs sector}
\label{radcorr}
The scalar sector of the $\mu\nu$SSM is subject to sizable
radiative corrections that have to be taken into account
in each phenomenologically viable analysis. Making these
corrections available to the public is (so far) the
core idea of this project.
The objects that contain the corrections are the
renormalized scalar self energies $\hat{\Sigma}_{\phi_i \phi_j}(p^2)$,
which enter the renormalized inverse propagator matrix
of the fields $\phi_i$,
\begin{equation}
\hat{\Gamma}_{ij} = \ii \left[
\delta_{ij}(p^2 - m_i^2) - \hat{\Sigma}_{\phi_i \phi_j}(p^2)
\right] \ .
\label{invpro}
\end{equation}
In this expression the indices $i$
and $j$ run over the number of
fields that mix with each other, $p$ is the
momentum and $m_i^2$ are the eigenvalues of
the corresponding tree-level
mass matrix. Implemented in the code are the
corrections to the CP-even and CP-odd neutral
scalars $h_i$ and $A_i$. These are given by
\begin{align}
\hat{\Sigma}_{h_i h_j} &= \hat{\Sigma}_{h_i h_j}^{(1)}(p^2) +
    \hat{\Sigma}_{h_i h_j}^{(2')} + \hat{\Sigma}_{h_i h_j}^{\mathrm{resum.}} \\
\hat{\Sigma}_{A_i A_j} &= \hat{\Sigma}_{A_i A_j}^{(1)}(p^2) \ .
\end{align}
Here, $\hat{\Sigma}_{h_i h_j}^{(1)}$ and
$\hat{\Sigma}_{A_i A_j}^{(1)}$ contain
the full one-loop corrections, including the
momentum dependence. In addition, leading
two-loop corrections for the CP-even fields
$h_i$ are included in terms of $\hat{\Sigma}_{h_i h_j}^{(2')}$.
Finally, higher-order corrections arising from
the resummation of logarithmic contributions
are taken into account in $\hat{\Sigma}_{h_i h_j}^{\mathrm{resum.}}$.
The corrections beyond one-loop level are taken from
the public code \texttt{FeynHiggs}. They are crucial
to obtain a precise prediction for the SM-like
Higgs-boson mass. $\hat{\Sigma}_{h_i h_j}^{(2')}$ contains
the fixed-order corrections of $\mathcal{O}(\alpha_t\alpha_s,
\alpha_b\alpha_s,\alpha_t^2,\alpha_t\alpha_b,\alpha_b^2)$ in the
approximation of vanishing electroweak gauge couplings
and ${p^2 = 0}$. $\hat{\Sigma}_{h_i h_j}^{\mathrm{resum.}}$
contains terms from the full resummation of leading and
next-to-leading logarithms and next-to-next-to-leading
logarithms of $\mathcal{O}(\alpha_s\alpha_t)$, obtained
from an effective theory calculation~\cite{Bahl:2018qog}.

The one-loop pieces were calculated
in \citere{Biekotter:2019gtq} in a mixed \DRbar-On Shell (OS) scheme that is
consistent with the one of \texttt{FeynHiggs}.
For generic scalar fields $\phi_i$, they can be written as
\begin{align}
\hat \Sigma_{\phi_i \phi_j}^{(1)}(p^2) &=
  \Sigma_{\phi_i \phi_j}^{(1)}(p^2)
  + \frac{1}{2} p^2
      \left( \delta Z_{\phi_j \phi_i} +
        \delta Z_{\phi_i \phi_j} \right) \notag \\
  &- \frac{1}{2}
      \left( m_{\phi_k \phi_j}^2
               \delta Z_{\phi_k \phi_i} +
             m_{\phi_i \phi_k}^2 \delta Z_{\phi_k \phi_j} \right)
  - \delta m_{\phi_i \phi_j}^2 \ .
\label{renorenselfsca}
\end{align}
$\Sigma_{\phi_i \phi_j}^{(1)}$ denotes the unrenormalized
self energies, extracted from the one-particle irreducible
scalar two-point functions. The field-renormalization
counterterms $\delta Z_{\phi_j \phi_i}$ and the mass counterterms
$\delta m_{\phi_i \phi_j}^2$ are defined in a way to cancel all
ultraviolet divergences appearing in $\Sigma_{\phi_i \phi_j}^{(1)}$.
The finite pieces of the counterterms are defined by the chosen
renormalization scheme. The field-renormalization
constants are defined as \DRbar\ parameters, such that
they do not contain finite terms. The mass
counterterms, on the other hand, are defined in a mixed
OS-\DRbar\ scheme. The gauge-boson masses $M_W$ and $M_Z$ and the
tadpole coefficients are renormalized applying
OS conditions, such that $\delta m_{\phi_i \phi_j}^2$
contains finite contributions
from the corresponding
counterterms~\cite{Biekotter:2019gtq}.

Without going into too much detail,
we summarize the numerical impact
of the radiative corrections on the Higgs-boson
masses of the $\mu\nu$SSM in the following.
Schematically, a rough approximation of the SM-like Higgs-boson
mass is given by
\begin{equation}
m_{h^{\rm SM}}^2 \sim
M_Z^2 \cos^2(2\beta) + \frac{1}{2}
    \lambda_i \lambda_i v^2 \sin^2(2\beta)
+ \Delta^{\rm MSSM}_{\rm (s)top} +
\Delta^{\widetilde{\nu}_{iR}^\mathcal{R}}_{\lambda_i^2} \ ,
\end{equation}
where the second term provides the enhancement of
the tree-level contribution compared to the
MSSM mentioned in
\refse{intro}. The third term consists of the
usual MSSM-like corrections from the stop and the top sector
(see \citere{Draper:2016pys} for a review).
In the gauge basis, these terms are practically unchanged in the
$\mu\nu$SSM. However, the mixing with the right-handed
sneutrinos modifies how much of $\Delta^{\rm MSSM}_{\rm (s)top}$ is finally
attributed to the mass eigenstate of the SM-like
Higgs boson.
The last term mainly arises from the mixing of the
doublet fields with the right-handed sneutrinos.
It was already observed in the next-to MSSM (NMSSM)
that, in contrast to the tree-level term
dependent on $\lambda_i$,
the loop-corrections contained in
$\Delta^{\widetilde{\nu}_{iR}^\mathcal{R}}_{\lambda_i^2}$
are usually negative and can, depending on the size of the
mixing, the value of $\tan\beta$ and
the self couplings $\kappa_{ijk}$, substantially
decrease the prediction for the SM Higgs-boson
mass~\cite{Ellwanger:2009dp}.
Due to the presence of three gauge singlet scalars
in the $\mu\nu$SSM instead of only one in the NMSSM,
the analytic form of 
$\Delta^{\widetilde{\nu}_{iR}^\mathcal{R}}_{\lambda_i^2}$
is much more complicated. However, the numerical
analysis of such corrections has shown that it is
crucial to take into account independently the
contributions from all three right-handed sneutrino
for a precise prediction of the SM-like Higgs-boson
mass~\cite{Biekotter:2019gtp}.

The radiative corrections to the right-handed
sneutrinos themselves are sizable only
for small masses in the vicinity of $125\gev$
or below~\cite{Biekotter:2017xmf,Biekotter:2019gtq}.
Otherwise, the tree-level mass
is already a good estimate. This is due
to the fact that the right-handed sneutrinos
are gauge singlets and only couple to the
SM particle content via a mixing with the
Higgs doublet fields.
If such mixing exists, the corresponding
right-handed sneutrino acquires additional
contributions to its mass from
$\Delta^{\rm MSSM}_{\rm (s)top}$.

Finally, the most interesting radiative corrections
are the ones obtained by the left-handed sneutrinos.
They are caused by genuine effects of the
$\mu\nu$SSM without a correspondence in the
(N)MSSM. It was shown that the dominant contributions
arise from the counterterms of the tadpoles, which
enter the mass counterterm in \refeq{renorenselfsca} with an
inverse factor of the vev of the scalar field under
consideration~\cite{Biekotter:2017xmf}.
For the left-handed sneutrinos this
means that they are enhanced by  the inverse of the small values
of $v_{iL}$. This enhancement can compensate the
suppression of factors of $Y^\nu_{ij}$ present in
lepton-number violating couplings. Here,
the corrections are mainly given by the tadpole
diagrams with the stops in the loop. The stops
are coupled to the left-handed sneutrinos
via an $F$-term tree-level coupling between
$\widetilde{t}_L$, $\widetilde{t}_R$,
$\widetilde{\nu}_{iL}$ and $\widetilde{\nu}_{iR}$,
after replacing $\widetilde{\nu}_{iR}$ with
the corresponding vev $v_{iR}$.
Expanding the complete renormalized self energy
in powers of $A^u_3 = A_t$ and $1/v_{iL}$, one
finds the very good approximation
\begin{align}
\hat{\Sigma}_{\vil \vil}^{(1)} &\approx
\left. \hat{\Sigma}_{\vil \vil}^{(1)} \right|_{\frac{\at}{v_{iL}}}
= \notag \\
&\frac{3}{16 \pi^2} \frac{v_u v_{iR}}{\sqrt{2} v_{iL}}
A_t Y_t^2 \yv \left( \frac{\log{(\frac{\mst}{\mu_R^2})}\mst -
	\log{(\frac{\mstt}{\mu_R^2})}\mstt}{\mst - \mstt} - 1 \right) \ ,
\label{corrsneul}
\end{align}
where $\mst$ and $\mstt$ are the squared stop masses and
$\mu_R$ is the renormalization scale.
These terms have to be added to the tree-level mass, which
is approximately given by
\begin{equation}
\label{eq:treeapp}
\left( m_{\vil \vil}^{(0)} \right)^2 \approx
\frac{\yv v_u v_{iR}}{\sqrt{2} v_{iL}}
\left( -\frac{1}{\sqrt{2}} \kappa_{iii} v_{iR} - \av \right) \ .
\end{equation}
This expression is subject to a renormalization-scale
dependence induced by the scale dependence of the \DRbar\
parameters. The numerically most sizable contribution
can be formulated approximately by the scale dependence of
$\av$, whose dominant piece is given by
\begin{equation}
\av(\mu_R,\mu_0) \approx
\av(\mu_0) +
	\frac{3}{16\pi^2}\yt^2\at\log\frac{\mu_R^2}{\mu_0^2} \ ,
\label{avscdep}
\end{equation}
that can be extracted from the \DRbar\ counterterm of
$\av$ as given in \citere{Biekotter:2019gtq}, and
where $\mu_0$ is the scale at which the value of $\av$
is given initially.
Combining all this, we find that the one-loop mass is given by
\begin{align}
\left( m_{\vil \vil}^{(1)} \right)^2 &\approx
\frac{\yv v_u v_{iR}}{\sqrt{2} v_{iL}}
\Bigg( -\frac{1}{\sqrt{2}} \kappa_{iii} v_{iR}- \av(\mu_0)
\notag \\
&\qquad\qquad-\frac{3}{16 \pi^2} \at \yt^2
	\Bigg(
	\frac{\log{\big(\frac{\mst}{\mu_0^2}\big)}\mst
	 - \log{\big(\frac{\mstt}{\mu_0^2}}\big)\mstt}
		{\mst - \mstt} - 1 \Bigg) \Bigg) \ .
\label{apprxmu0}
\end{align}
The logarithmic terms can be further simplified
under the assumption that
\begin{equation}
\mst - \mstt \ll M_{S}^2 \approx \mst \approx \mstt \ ,
\end{equation}
with $M_S$ being the \textsc{Susy}-breaking scale,
such that
\begin{align}
\left( m_{\vil \vil}^{(1)} \right)^2 &\approx
\frac{\yv v_u v_{iR}}{\sqrt{2} v_{iL}}
\Bigg( -\frac{1}{\sqrt{2}} \kappa_{iii} v_{iR}- \av(\mu_0)
-\frac{3}{16 \pi^2} \at \yt^2
	\log\big( \frac{M_S^2}{\mu_0^2} \big) \Bigg) \ .
\end{align}
Note that the renormalization-scale dependence
of the radiative corrections given in \refeq{corrsneul}
drops out once the scale dependence
of $\av$ is considered. Instead, the size of the
corrections depends on the input scale of the
\DRbar\ parameters $\mu_0$. The corrections
vanish if $\mu_0$ is chosen to be close to the
stop masses. Furthermore, it is convenient to
choose the renormalization scale $\mu_R$ to be
equal to the input scale $\mu_0$, so that the
logarithmic term in \refeq{avscdep} vanishes,
and the tree-level expectation for the left-handed
sneutrino mass given in \refeq{eq:treeapp} is unchanged.
This is why in the code presented here the scales
are fixed by default to be
\begin{equation}
\mu_0 = \mu_R = M_S \ ,
\end{equation}
such that
\begin{equation}
\left( m_{\vil \vil}^{(1)} \right)^2 \approx
\left( m_{\vil \vil}^{(0)} \right)^2 \approx
\frac{\yv v_u v_{iR}}{\sqrt{2} v_{iL}}
\left( -\frac{1}{\sqrt{2}} \kappa_{iii} v_{iR} -
    \av(M_S) \right) \ .
\end{equation}
Even though in principle any choice for the
scales would be equally valid
(within a physically reasonable range), the choice given
above is highly recommended as long as the calculation of
radiative corrections to the slepton masses has not been
carried out. The reason is that large loop corrections
to the left-handed sneutrinos could artificially
change the mass ordering of the left-handed sneutrinos
and sleptons, just because they are treated at
different orders of perturbation theory, and
therefore modify the phenomenology of a benchmark
point completely. However,
due to the different $D$-term contributions it is known
that a left-handed sneutrino of a certain flavour
cannot be heavier than the corresponding left-handed slepton,
such that these artificial effects are unphysical
and must be avoided.

\section{The python package \texttt{munuSSM}}
\label{secmunu}
\begin{figure}
\centering
\includegraphics[height=0.8\textheight]{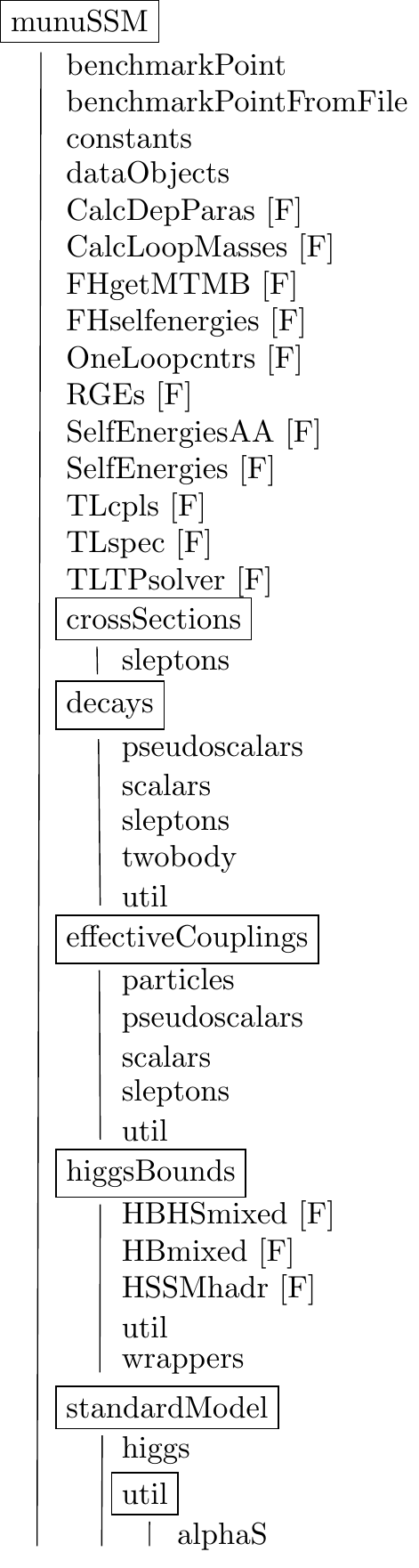}
\caption{General structure of the code. Packages are indicated
with the squared boxes, and the vertical lines indicate to which
package each module belongs. Modules written in Fortran are marked
with an [F].}
\label{genstruct}
\end{figure}

In this section we present the general structure
of the code, which is also depicted in \reffi{genstruct}.
The main package is called \texttt{munuSSM} and it
contains the subpackages \texttt{crossSections},
\texttt{decays}, \texttt{effectiveCouplings},
\texttt{higgsBounds} and \texttt{standardModel}.
Note that some of the modules are written in Fortran
and compiled to python libraries using the compiler
\texttt{f2py} from \texttt{NumPy}~\cite{Harris:2020xlr}.

The usage of Fortran
has several advantages. Firstly, numerical calculations are
much faster in a statically typed language like Fortran.
In addition, the numerical precision of floating-point
numbers can be enhanced to quadruple precision in Fortran.
In the context of the $\mu\nu$SSM, this turned out to
be necessary due to the hierarchical structure of particle
masses and mixing patterns. In particular, the seesaw
mechanism leads to a mass matrix
for the neutral fermions
whose eigenvalues
range from sub-eV to the TeV values, which
is numerically challenging to diagonalize.
Finally, the codes that are
interfaced are all written in Fortran, such that it is
much easier to use their libraries within a Fortran routine
that is subsequently compiled to python.
For the user of the package \texttt{munuSSM} the usage
of Fortran is largely irrelevant. The only thing that is
important is that the parameters of the model are not saved
as usual python float objects and \texttt{NumPy} arrays,
but as \texttt{numberQP} and \texttt{arrayQP} objects, that
are defined in the module \texttt{dataObjects}. To obtain
the values as floats or float arrays, the user just has
to type \texttt{a.float} in case of \texttt{a} being
an instance of \texttt{numberQP} or \texttt{arrayQP}.

The user interface is defined in
the class \texttt{BenchmarkPointFromFile}. This class inherits the
methods of the class \texttt{BenchmarkPoint} to construct
and analyze a benchmark point. In addition, it reads
the input parameters from a file during the initialization.
Within this class, the subpackages are utilized to calculate
the branching ratios and cross sections of the scalar particles,
while the modules of the main package \texttt{munuSSM} perform
the calculation of the radiatively corrected particle spectrum.
Before explaining the methods of the \texttt{BenchmarkPoint}
class and how the user can call them, we briefly explain the
role of each subpackage and give some details on the implementation.

\begin{table}
\hspace*{-.8cm}
{\renewcommand{\arraystretch}{1.2}
\begin{tabular}{p{0.44\textwidth}p{0.56\textwidth}}
\textbf{BenchmarkPoint} & \\
\hline
\vspace{-1.8em}
\begin{verbatim}
calc_tree_level_spectrum(
    self)
\end{verbatim} &
Calculates the particle spectrum at tree level. \\
\vspace{-3.1em}
\begin{verbatim}
calc_tree_level_couplings(
    self)
\end{verbatim} &
\vspace{-2.4em}
Calculates the complete set of couplings at tree level. \\
\vspace{-3.1em}
\begin{verbatim}
calc_one_loop_counterterms(
    self)
\end{verbatim} &
\vspace{-2.4em}
Calculates the counterterms used in the renormalized one-loop self energies. \\
\vspace{-2.0em}
\begin{verbatim}
calc_one_loop_self_energies(
    self,
    even,
    odd,
    p2_Re,
    p2_Im)
\end{verbatim} &
\vspace{-1.2em}
Calculates the values of the renormalized one-loop self energies
for the CP-even scalars if \texttt{even=1} and for the CP-odd
scalars if \texttt{odd=1} for a given momentum
$p$, where $\texttt{p2\_Re}=\mathrm{Re}(p^2)$
and $\texttt{p2\_Im}=\mathrm{Im}(p^2)$. \\
\vspace{-2.7em}
\begin{verbatim}
calc_two_loop_self_energies(
    self,
    p2_Re,
    p2_Im)
\end{verbatim} &
\vspace{-2.0em}
Calculates the values of the renormalized self energies
with the full one-loop and partial higher-order corrections
for the CP-even scalars, with \texttt{p2\_Re} and \texttt{p2\_Im}
as defined before. \\
\vspace{-1.6em}
\begin{verbatim}
calc_loop_masses(
    self,
    even=2,
    odd=1,
    accu=1.e-5,
    momentum_mode=1)
\end{verbatim} &
\vspace{-1.0em}
Calculates the loop corrected scalar masses.
\texttt{even=0,1,2} selects
the loop level for the CP-even scalars
(2 includes all higher-order corrections).
\texttt{odd=0,1} selects the
loop level for the CP-odd scalars.
\texttt{momentum\_mode=0,1} selects the
treatment of finite momenta (see text).
\texttt{accu} is the numerical precision of the matrix diagonalization. \\
\vspace{-1.7em}
\begin{verbatim}
calc_effective_couplings(
    self)
\end{verbatim} &
\vspace{-1.0em}
Calculates the effective couplings of the CP-even
and the CP-odd scalars \\
\vspace{-2.8em}
\begin{verbatim}
calc_branching_ratios(
    self)
\end{verbatim} &
\vspace{-2.2em}
Calculates the decay widths and branching ratios of the CP-even,
the CP-odd and the charged scalars.
\end{tabular}
}
\caption{Class methods of the class \texttt{BenchmarkPoint} as
defined in the module \texttt{benchmarkPoint}.}
\label{bpmeth}
\end{table}

\begin{itemize}
\item[-] \texttt{crossSections} contains the calculation of
cross sections of particles at the LHC or any other future collider.
So far the only cross section implemented is the charged Higgs-boson
production in the $p p \rightarrow H^\pm t b$ channel, which is however
only relevant for the charged scalar of the $\mu\nu$SSM corresponding
to the charged Higgs boson of the MSSM. For the remaining sleptons,
the couplings to quarks are suppressed by the smallness of lepton-number
violation, as is their mixing with the MSSM-like
charged Higgs boson. The above mentioned cross section is implemented
in the form of a spline interpolation as a function of
$\tan\beta$ and the charged Higgs-boson mass
in the 2HDM limit~\cite{deFlorian:2016spz},
therefore lacking subdominant \textsc{Susy}-QCD
corrections. For the charged scalars corresponding to the sleptons,
the main production channel is the production in pairs, which is
currently not yet implemented.\footnote{This is partially due to
the fact that the pair-production cross sections of charged Higgs
bosons is currently unused within \texttt{HiggsBounds}, even though
it can be given as input~\cite{Bechtle:2020pkv}.}
The cross sections of the neutral scalars are obtained via the
interface to \texttt{HiggsBounds}, based on the effective couplings
calculated in the supackage \texttt{effectiveCouplings}
(see below).
\item[-] \texttt{decays} calculates the decay widths and branching ratios
of all neutral and charged Higgs bosons. The decay widths of decays
into SM particles are implemented via a rescaling of the SM prediction
for a Higgs boson of the same mass, again utilizing the effective
couplings calculated in the \texttt{effectiveCouplings} package.
The SM predictions are implemented in the form of
cubic spline interpolations of data tables
published in \citeres{Heinemeyer:2013tqa,deFlorian:2016spz},
based on the results obtained with the codes
\texttt{HDECAY}~\cite{Djouadi:1997yw,Spira:1997dg,Butterworth:2010ym}
and \texttt{PROPHECY4F}~\cite{Bredenstein:2006rh,Bredenstein:2006ha}.
The decays into BSM particle final states are considered at
leading order, however using the Higgs-boson couplings rotated
into the radiatively corrected mass eigenstate basis, therefore
taking into account the propagator corrections calculated in the
main package \texttt{munuSSM}. The implementation of these decays
follows the general approach of \citere{Goodsell:2017pdq}.
Using the couplings in the loop-corrected basis corresponds to
taking into account the finite wave-function renormalization factors
(or $Z$-factors) in the limit of vanishing
momentum~\cite{Frank:2006yh,Goodsell:2017pdq}.
For the accurate prediction of the Higgs-boson masses, it is
recommended to include the momentum dependence of the radiative
corrections. Strictly speaking, this leads to the fact that the
mixing matrices will not be unitary anymore. Fortunately, these
effects are numerically negligible except for extreme cases.
So far, the only three-body decays considered are the decays
into off-shell vector bosons, whose corresponding decay widths
are included in the SM prediction for the decays into a
pair of massive gauge bosons.
\item[-] \texttt{effectiveCouplings} calculates the effective
couplings of the neutral scalars, defined as the coupling strength
normalized to the one of a hypothetical
SM Higgs boson having the same mass. The precise definition
of these coefficients can be found in \citere{Bechtle:2020pkv}.
Loop-induced couplings, as the ones to photons or gluons,
are calculated using the general expressions
for the form factors as can be found in
\citere{Spira:2016ztx}.
Resummed higher-order corrections proportional to $\tan\beta$
are implemented for couplings to the third generation of
down-type fermions in terms of the quantities $\Delta_b$
and $\Delta_\tau$ following \citere{Spira:2016ztx}.
As already mentioned before, the effective couplings
are used to calculate decays into SM particles. Apart from that,
they are given as input to the code \texttt{HiggsBounds}, which
uses them to calculate the production cross
sections at LEP, Tevatron and the LHC.\footnote{In the
traditional effective-coupling input
of \texttt{HiggsBounds}, the effective couplings are also used
internally to calculate branching ratios. In our
interface, we use a mixed input in which the branching ratios
are given as additional input as calculated in the subpackage
\texttt{decays}. Effectively, this corresponds to the SLHA input
format of \texttt{HiggsBounds}.}
\item[-] \texttt{higgsBounds} constructs the input arrays
for the interface to \texttt{HiggsBounds} and \texttt{HiggsSignals}.
In addition, it provides a wrapper class to directly call both codes
from within python.\footnote{A stand-alone python wrapper for
\texttt{HiggsBounds} can be found under
\url{https://gitlab.com/thomas.biekoetter/higgsbounds_python_wrapper}.}
Since we interact with both external codes
via their Fortran libraries, we can save additional results
beyond the usual output, such as cross sections. Via the
\texttt{higgsBounds} subpackage, a given set of benchmark points of
the $\mu\nu$SSM can easily be tested against constraints from
collider searches and the signal rates of the SM-like Higgs boson.
\item[-] \texttt{standardModel} contains the data tables of the
SM predictions for decay widths of the Higgs boson as given
in \citeres{Heinemeyer:2013tqa,deFlorian:2016spz}. The data is
given for different mass intervals.
The subpackage constructs
spline interpolations of the data and provides functions
taking the Higgs-boson mass as input to
extract the data.
The maximum value for the particle mass of the data tables is
at around $1\tev$. If within the \texttt{decay} subpackage
larger masses appear, the values are extrapolated based
on the known leading mass
dependence~\cite{Spira:2016ztx}.
\end{itemize}

\noindent Having explained the role of the
subpackages, we now turn to the main package \texttt{munuSSM}.
Therein, basically all calculations are performed within
Fortran modules. During the initialization of a
benchmark point, the Fortran modules
\texttt{CalcDepParas} and \texttt{TLTPsolver}
set up the complete set of model parameters.
The latter solves the tadpole equations for the
diagnoal soft mass parameters given the vevs
as input. The module \texttt{FHgetMTMB} calls
\texttt{FeynHiggs} to extract
the top-quark and bottom-quark masses used in the
scalar self energies (see \refse{secfh}
for details). The tree-level spectrum and the
couplings are calculated in the modules
\texttt{TLspec} and \texttt{TLcpls}. These are
then used for the calculation of the renormalized
self energies. They are implemented in the form
as shown in \refeq{renorenselfsca}. The counterterms are
independent of the momentum, so that they are
calculated only once in the module \texttt{OneLoopcntrs}.
Once they are available, the one-loop part of the self energies of
the CP-even and the CP-odd scalars are calculated
in the modules \texttt{SelfEnergies} and \texttt{SelfEnergiesAA}.
The contributions beyond one-loop level are
extracted from \texttt{FeynHiggs} in the
module \texttt{FHselfenergies}. Finally,
the loop-corrected scalar spectrum is
calculated in the module \texttt{CalcLoopMasses} by finding
the zeros of the determinant of the inverse
propagator matrix shown in \refeq{invpro}.

The user interface is defined via the methods of the python class
\texttt{BenchmarkPoint}, such that the Fortran modules described
before do not have to be called directly by the user.
The complete list of public routines
is listed in \refta{bpmeth}. Note that an instance of this class
should be created via its subclass \texttt{BenchmarkPointFromFile},
which contains additional routines to read the parameter values
from an input file.
Because of potentially large corrections to the
masses of the left-handed sneutrinos (see \refse{radcorr}),
the renormalization scale $\mu_R$ at which the
radiative corrections are evaluated is set to
be equal to the \textsc{Susy}-breaking scale $M_S$ at
which the \DRbar\ \textsc{Susy} parameters are defined.
While these are given as input by the user, the SM parameters
are set to default values in the module \texttt{constants}.
In this module, also the value for $M_S$ is fixed to
$1\tev$ by default. This value should only be changed by the
user if the stop masses are much heavier than $1\tev$.
Note, however, that in such a situation the Feynman-diagrammatic
fixed-order calculation applied in this code is not the
most accurate one and a hybrid approach (as is implemented
in \texttt{FeynHiggs}) incorporating
effective field theory calculations is required.

For a phenomenological study of a benchmark point, the
most interesting routines for the user are
\texttt{calc\_loop\_masses} to obtain a precise
prediction for the particle spectrum and
\texttt{calc\_branching\_ratios} to obtain the
branching ratios of the scalars.
The remaining functions can be called directly by
the user, but will usually be called only internally,
as they provide the required quantities for the above
mentioned functions. For instance, if the user calls
\begin{lstlisting}
pt.calc_loop_masses(2, 1, momentum_mode=1)
\end{lstlisting}
with \texttt{pt} being an instance of
\texttt{BenchmarkPointFromFile}, it is internally checked
if the tree-level masses and couplings are already available.
If they are not, the functions \texttt{calc\_tree\_level\_spectrum}
and \texttt{calc\_tree\_level\_couplings} are automatically called before
calculating the radiative corrections.
With \texttt{momentum\_mode=1} we choose to take into account
the momentum dependence of the radiative corrections.
\texttt{momentum\_mode=0} selects the limit of vanishing
external momentum. This options is less precise but faster, because the
inverse propagator matrix has to be diagonalized
only once, while an iterative procedure is applied for
\texttt{momentum\_mode=1}.
In the same manner, if
\begin{lstlisting}
pt.calc_branching_ratios()
\end{lstlisting}
is called, the effective couplings are required for the
rescaling of the SM predictions, such that internally
\texttt{calc\_effective\_couplings} is called if it
has not already been called before.
In \ref{return} we state the exact form of the return
values and class attributes set by each function shown
in \refta{bpmeth}. Basic user instructions are given
in \refse{usage}. Before that we provide some details
on the interfaces to the other public codes.

\subsection{Interfaces}
\label{dringesichert}
The package \texttt{munuSSM} makes use of other
public codes for some of the model predictions.
This codes are downloaded
and installed automatically during the installation of the
main package (see \refse{insta}).
The interfaces utilize the Fortran
libraries of the codes. In the following we briefly 
describe the information provided by the codes
and how they are called internally.

\subsubsection{FeynHiggs}
\label{secfh}
For the accurate prediction of the SM-like Higgs-boson
mass, a pure one-loop calculation is not sufficient.
Fortunately, the dominant higher-order corrections can
be taken over in approximate form
from the MSSM. However, one has to take care
of a consistent combination of the one-loop corrections
calculated in the full $\mu\nu$SSM and the higher-order
corrections known from the MSSM.
This is why in \citeres{Biekotter:2017xmf,Biekotter:2019gtq}
the renormalization prescription of the one-loop
calculation in the $\mu\nu$SSM was closely based on the
one implemented in the public MSSM code \texttt{FeynHiggs},
such that the higher-order corrections could be supplement
from there.

The radiative corrections to the scalar masses and mixings
are given by the renormalized self energies $\hat\Sigma(p^2)$
that enter the inverse propagator matrix as shown in
\refeq{invpro}. Schematically, the self energies of the
CP-even Higgs bosons are implemented as
\begin{equation}
\hat\Sigma = \hat\Sigma^{(1)}_{\mu\nu\rm SSM} -
\hat\Sigma^{(1)}_{\rm FeynHiggs} +
\hat\Sigma^{(1)+(2')+\rm resum.}_{\rm FeynHiggs} \ .
\label{schema}
\end{equation}
The piece $\hat\Sigma^{(1)}_{\mu\nu\rm SSM}$ is the
full one-loop result including all couplings of the
$\mu\nu$SSM, and renormalized according to \refeq{renorenselfsca}.
The numerical evaluation of the loop functions
appearing in $\hat\Sigma^{(1)}_{\mu\nu\rm SSM}$
is achieved via a link to the public code
\texttt{LoopTools}~\cite{Hahn:1998yk}.
Imaginary parts of the loop momentum $p^2$ are
considered via a Taylor expansion with respect
to $\mathrm{Im}(p^2)$ up to first order.
To this piece we add the full \texttt{FeynHiggs v.2.16.1}
result including the approximate two-loop contributions
and the contributions obtained from the resummation
of logarithmic contributions denoted by the
term $\hat\Sigma^{(1)+(2')+\rm resum.}_{\rm FeynHiggs}$.
Since this piece also contains the MSSM
one-loop result, these terms have to be subtracted again
to avoid a double counting. This is done by calling
\texttt{FeynHiggs} a second time with the flag
\texttt{looplevel} set to 1, yielding
$\hat\Sigma^{(1)}_{\rm FeynHiggs}$, which is
then subtracted from the sum.

For this procedure to be consistent, it is crucial
that the one-loop piece of the $\mu\nu$SSM is calculated
using the same set of parameters as is used in
\texttt{FeynHiggs}. In particular, this concerns the
values of the top-quark mass and the bottom-quark
mass, from which the corresponding Yukawa couplings
$Y_t$ and $Y_b$ are derived.\footnote{Note that the strong
QCD coupling constant $\alpha_s$ does not enter at one-loop level.}
This is achieved by a slightly modified version of the
\texttt{FeynHiggs} routine \texttt{FHGetPara}, which is
called during the initialization of an instance of
\texttt{BenchmarkPointFromFile}. For the top quark, the pole
mass $M_t$ is given as input and \texttt{FHGetPara} returns
the \MSbar\ value of the top-quark mass at the scale $M_t$ in the
SM at NNLO $\overline{m}_t^{\MSbar,\rm SM}(M_t)$, which is used in
\texttt{FeynHiggs} for the calculation of
$\hat\Sigma^{(1)+(2')+\rm resum.}_{\rm FeynHiggs}$. In principle, the
value is different when  the log resummation
is switched off with \texttt{loglevel=0},
such that the value of $\overline{m}_t^{\MSbar,\rm SM}(M_t)$
would be different in
$\hat\Sigma^{(1)}_{\rm FeynHiggs}$, yielding a mismatch
compared to $\hat\Sigma^{(1)}_{\mu\nu\rm SSM}$.
To avoid that, we set by hand
the flag \texttt{loglevelmt=3} in the
\texttt{FeynHiggs} routine \texttt{FHSetFlags}, so that
the same value of $\overline{m}_t^{\MSbar,\rm SM}(M_t)$
is used independently of the flag \texttt{loglevel}.

In a similar way, we obtain $\overline{m}_b^{\DRbar,\rm MSSM}(M_S)$,
i.e., the MSSM \DRbar-renormalized value of the
bottom-quark mass at the scale $M_S$, in the
modified routine \texttt{FHGetPara}.
We extract the value used by \texttt{FeynHiggs}
when called with \texttt{looplevel=1}. In contrast to
$\overline{m}_t^{\MSbar,\rm SM}(M_t)$,
which is given by SM RGEs, the precise value
of $\overline{m}_b^{\DRbar,\rm MSSM}(M_S)$
depends also on the \textsc{Susy} parameters, mainly
via the so-called $\Delta_b$-corrections.
Apart from that, it is different when called with
\texttt{looplevel=2}. However, for the prescription in \refeq{schema}
to be consistent, this is not a problem as long as we
assure that the value of the quark masses
in $\hat\Sigma^{(1)}_{\mu\nu\rm SSM}$
and $\hat\Sigma^{(1)}_{\rm FeynHiggs}$ are identical.

For the remaining MSSM one-loop contributions, arising from
loop diagrams with particles inserted in the loop that are not
(s)tops or (s)bottoms, the double-counting is automatically
avoided due to the cancellation between $\hat\Sigma^{(1)}_{\rm FeynHiggs}$
and $\hat\Sigma^{(1)+(2')+\rm resum.}_{\rm FeynHiggs}$, because
they do not depend on the flags \texttt{looplevel}
or \texttt{loglevel}.
Thus, only the one-loop result in the full model
contained in $\hat\Sigma^{(1)}_{\mu\nu\rm SSM}$
contributes for these sectors.
This is important because they might be substantially
modified compared to the MSSM. For instance, due to
the presence of the portal couplings $\lambda_i$,
the tree-level masses of the doublet-like Higgs bosons
receive additional contributions, so that loop diagrams
with Higgs bosons in the loop have to be accounted for
in $\hat\Sigma^{(1)}_{\mu\nu\rm SSM}$,
while the corresponding diagrams from the MSSM
should drop out.

Our approach using \texttt{FeynHiggs} does not capture
the modifications of the tree-level Higgs sector
of the $\mu\nu$SSM compared to the MSSM
proportional to $\lambda_i$ within the contributions
beyond one-loop level. They would enter in the approximate
two-loop result via the fixed-order terms of
$\mathcal{O}(\alpha_t^2,\alpha_b^2,\alpha_b\alpha_t)$,
in which the Higgs bosons appear as internal particles
in the corresponding loop diagrams. However,
this is a subleading effect as long as the corrections
to the doublet fields are dominant. Also, it is the
best possible approximation while the
calculation of the two-loop contributions in the full
model is not carried out. 
Nevertheless, for small values of $\tan\beta$ and
large values of $\lambda_i$ this leads to a potential
source of theory uncertainty for the prediction of
the SM-like Higgs-boson mass.
In comparison to neglecting the contributions
beyond one-loop level entirely,
our numerical results
of \citeres{Biekotter:2017xmf,Biekotter:2019gtq}
showed that even in these cases
the prediction for the Higgs-boson
mass improves when taking the approximate MSSM
contributions into account.
The same conclusion was drawn in other
analyses using \texttt{FeynHiggs} for
similar extensions of the
MSSM~\cite{Drechsel:2016htw,Hollik:2018yek,Hollik:2018bwj}.

Once the renormalized self energies are constructed, the
inverse propagator matrix is diagonalized using the
public Fortran library \texttt{Diag}~\cite{Hahn:2006hr}.
If the momentum dependence is taken into account, the
loop-corrected pole masses are given by the zeros of
the determinant of the inverse propagator matrix,
which are calculated by an iterative procedure.

\subsubsection{HiggsBounds}
\label{hbsec}

\begin{table}
{\renewcommand{\arraystretch}{1.2}
\begin{tabular}{p{0.14\textwidth}p{0.14\textwidth}p{0.7\textwidth}}
\textbf{self.HiggsBounds} & \\
\hline
\texttt{result} & \texttt{(23, )} &
The first element is the global \texttt{HiggsBounds}
result with \texttt{1=allowed} and \texttt{0=forbidden}.
The following 22 elements are the
results for each CP-even, CP-odd and
charged scalar, in this order with ascending masses. \\
\texttt{chan} & \texttt{(23, )} &
As before, but each element gives the channel number of the
most sensitive search, as listed in the file \texttt{Key.dat}
that \texttt{HiggsBounds} produces. \\
\texttt{obsratio} & \texttt{(23, )} &
As before, but each element gives the ratio of predicted and
observed channel rate. \\
\texttt{ncombined} & \texttt{(23, )} &
As before, but each element gives the number of scalars
whose signal rates were combined and assumed to be
contributing to the search channel. \\
\texttt{XSsingleH} & \texttt{(15, 4)} &
Hadronic inclusive single-Higgs production cross section at the
Tevatron with $2\tev$ and the LHC with
7, 8 and $13\tev$ center-of-mass energy for
each CP-even and CP-odd scalar,
normalized to the SM prediction. \\
\texttt{XSggH} & \texttt{(15, 4)} &
As before, but for gluon fusion production. \\
\texttt{XSbbH} & \texttt{(15, 4)} &
As before, but for $b \bar b$ associated production. \\
\texttt{XSVBF} & \texttt{(15, 4)} &
As before, but for vector boson fusion production. \\
\texttt{XSWH} & \texttt{(15, 4)} &
As before, but for production in association with
a $W$. \\
\texttt{XSZH} & \texttt{(15, 4)} &
As before, but for production in association with
a $Z$. \\
\texttt{XSttH} & \texttt{(15, 4)} &
As before, but for $t \bar t$ associated production. \\
\texttt{XStH\_tchan} & \texttt{(15, 4)} &
As before, but for single $t$ associated production
through $t$-channel exchange. \\
\texttt{XStH\_schan} & \texttt{(15, 4)} &
As before, but for single $t$ associated production
through $s$-channel exchange. \\
\texttt{XSqqZH} & \texttt{(15, 4)} &
As before, but for quark-initiated production in
association with a $Z$ boson. \\
\texttt{XSggZH} & \texttt{(15, 4)} &
As before, but for gluon-initiated production in
association with a $Z$ boson.
\end{tabular}
}
\caption{Form of the dictionary \texttt{HiggsBounds}
containing the results of the \texttt{HiggsBounds} routine
as set by the function \texttt{check\_higgsbounds}.
The first column lists the keys of the dictionary.
The items of each key are \texttt{NumPy} arrays with the
shape given in the second column. The third column
explains the meaning of each entry.}
\label{hbdict}
\end{table}

To test a set of benchmark points against collider
constraints from searches for BSM scalars, an
interface to the public code \texttt{HiggsBounds~v.~5.9.0}
is implemented. With \texttt{pts} being a single instance or
a list of instances of the class
\texttt{BenchmarkPoint}, the user can call the
function
\begin{lstlisting}
check_higgsbounds(pts)
\end{lstlisting}
defined in the module \texttt{util} of the
subpackage \texttt{higgsBounds}.
This function first calls the method
\texttt{\_setup\_higgsbounds} for each instance of
\texttt{BenchmarkPoint} in \texttt{pts}, which will subsequently
call \texttt{calc\_effective\_couplings} and
\texttt{calc\_branching\_ratios} (see \refta{bpmeth})
in case they have not been called before.
Based on the effective couplings and the branching
ratios, \texttt{check\_higgsbounds} will then construct
the input for \texttt{HiggsBounds} for the whole
set of points. Finally, the \texttt{HiggsBounds} library
is accessed
via the Fortran module \texttt{HBmixed}.
This module is called
within the wrapper class \texttt{Mixed} defined in
the subpackage \texttt{higgsBounds}.

The results are saved as dictionaries which are
set as class attributes to each benchmark point contained
in \texttt{pts}. If \texttt{pt} is an instance of
\texttt{BenchmarkPoint}, the results are saved in:
\begin{lstlisting}
pt.HiggsBounds
\end{lstlisting}
This dictionary has the elements listed in \refta{hbdict}.
The meaning of each entry of
the dictionary corresponds to the original definitions
within \texttt{HiggsBounds}~\cite{Bechtle:2020pkv}.
The user can check if the benchmark point is excluded
depending on the value:
\begin{lstlisting}
pt.HiggsBounds['result'][0]
\end{lstlisting}
It is 1 if the point is allowed and 0 if any of
the scalars is excluded. With the remaining elements
of this array,
the user can verify which of the scalars are excluded.
The experimental search responsible for the exclusion
can be obtained by comparing the channel number saved
under the key \texttt{chan} with the list of applied
experimental searches saved in the
file \texttt{Key.dat} that \texttt{HiggsBounds}
creates automatically. The cross sections
for the neutral scalars that are calculated by
\texttt{HiggsBounds} rely on the effective couplings
calculated before.

\subsubsection{HiggsSignals}
In addition to the test against cross-section limits
using \texttt{HiggsBounds}, it is possible to verify
whether a benchmark point contains a Higgs boson at $\sim 125\gev$
that correctly accommodates the measured signal rates
of the SM-like Higgs boson. This is done via an interface
to the public code \texttt{HiggsSignals~v.~2.5.1}. Since \texttt{HiggsSignals}
relies on the \texttt{HiggsBounds} subroutines to read the
theoretical input, it is reasonable to combine both tests
into a single function call. We provide the function
\begin{lstlisting}
check_higgsbounds_higgssignals(pts)
\end{lstlisting}
defined in the module \texttt{util} of the subpackage \texttt{higgsBounds}.
As before, \texttt{pts} can be a single instance or a list
of instances of the class \texttt{BenchmarkPoint}.
Executing the above command will call both \texttt{HiggsBounds}
and \texttt{HiggsSignals} via the Fortran module
\texttt{HBHSmixed}. For a better interpretation of the
$\chi^2$ test performed by \texttt{HiggsSignals}, \texttt{HiggsSignals} is
called a second time via the Fortran module \texttt{HSSMhadr},
providing a reference $\chi^2_{\rm SM}$ value based on the SM
predictions using the same set of experimental measurements.

\begin{table}
\centering
{\renewcommand{\arraystretch}{1.4}
\begin{tabular}{p{0.22\textwidth}p{0.7\textwidth}}
\textbf{self.HiggsSignals} & \\
\hline
\texttt{Chisq\_mu} &
The $\chi^2_\mu$ result regarding the signal-rate measurements. \\
\texttt{Chisq\_mh} &
The $\chi^2_{m_h}$ result regarding the mass measurements. \\
\texttt{Chisq} &
The total $\chi^2$ result, i.e., $\chi^2 = \chi^2_\mu + \chi^2_{m_h}$. \\
\texttt{nobs} &
The total number of observables considered in the $\chi^2$ test. \\
\texttt{Pvalue} &
The $p$ value derived from the $\chi^2$ result assuming one
free parameter. \\
\texttt{Delta\_Chisq\_mu} &
The difference $\chi^2_\mu - \chi^2_{\mu, \rm SM}$, where
$\chi^2_{\mu, \rm SM}$ is the Standard Model reference $\chi^2$
evaluated using the same set of signal-rate measurements. \\
\texttt{Delta\_Chisq\_mh} &
As before, but for the difference $\chi^2_{m_h} - \chi^2_{m_h, \rm SM}$
using the same set of mass measurements. \\
\texttt{Delta\_Chisq} &
As before, but for the difference $\chi^2 - \chi^2_{\rm SM}$, where
$\chi^2_{\rm SM} = \chi^2_{\mu, \rm SM} + \chi^2_{m_h, \rm SM}$.
\end{tabular}
}
\caption{Form of the dictionary \texttt{HiggsSignals}
containing the results of the \texttt{HiggsSignals} routine
as set by the function \texttt{check\_higgsbounds\_higgssignals}.
The first column lists the keys of the dictionary.
The second column explains the meaning of each entry.}
\label{hsdict}
\end{table}

The complete result of the function
\texttt{check\_higgsbounds\_higgssignals} is saved
as dictionaries in the class attributes:
\begin{lstlisting}
pt.HiggsBounds
pt.HiggsSignals
\end{lstlisting}
As already mentioned before, \texttt{pt} is an instance of the class
\texttt{BenchmarkPoint} contained in \texttt{pts}.
The dictionary \texttt{pt.HiggsBounds} was already
introduced in \refse{hbsec} (see also \refta{hbdict}).
The dictionary \texttt{pt.HiggsSignals} contains the
\texttt{HiggsSignals} results. The whole list of
entries is given in \refta{hsdict}.
For the interpretation of the fit, the most valuable
information is provided by the global $\chi^2$ value
contained in
\begin{lstlisting}
pt.HiggsSignals['Chisq']
\end{lstlisting}
and the difference of this value to the SM reference
value contained in:
\begin{lstlisting}
pt.HiggsSignals['Delta_Chisq']
\end{lstlisting}
We leave it to the user to decide
which values are considered to represent an accurate
fit to the experimental data. For more information
about the interpretation of the \texttt{HiggsSignals}
results we refer to \citere{Bechtle:2014ewa}.
We recommend to define a criteria based on
the difference between the $\chi^2$ value and the SM
reference value $\chi^2_{\rm SM}$, instead of only
taking into account the $\chi^2$ value of the benchmark
point alone.

\subsection{Installation}
\label{insta}
To install the package \texttt{munuSSM} you
need the version control system \texttt{git},
working compilers for Fortran, c and c++
(recommended \texttt{gfortran} and \texttt{gcc}),
and \texttt{cmake} for the installation
of \texttt{HiggsBounds} and \texttt{HiggsSignals}.
All of this is already installed on a regular unix
machine. You can clone the repository with
\texttt{SSH} by typing:
\begin{verbatim}
    git clone git@gitlab.com:thomas.biekoetter/munussm.git
\end{verbatim}
Alternatively, you can clone the repository
with \texttt{HTTPS} by typing:
\begin{verbatim}
    git clone https://gitlab.com/thomas.biekoetter/munussm.git
\end{verbatim}
Then the package can be installed by entering the directory and
executing the makefile:
\begin{verbatim}
    cd munussm
    make all
\end{verbatim}
You can specify the python version used for the
installation by typing, for instance:
\begin{verbatim}
    make all PC=python3.6
\end{verbatim}
We stress that python version 2 is not supported.
Furthermore, if you wish to specify the \texttt{gnu}
compiler versions, you can type, for instance:
\begin{verbatim}
    make all FC=gfrotran-10 CC=gcc-10 CXX=g++-10
\end{verbatim}
During the installation process, the external
libraries \texttt{Diag}, \texttt{LoopTools},
\texttt{FeynHiggs}, \texttt{HiggsBounds} and
\texttt{HiggsSignals} are installed in the
directory \texttt{external}.
Once the installation process terminated, the
package is installed in your python environment
and can be imported with:
\begin{verbatim}
    import munuSSM
\end{verbatim}

\subsection{Usage}
\label{usage}
Only basic knowledge of the python programming
language is required to use the package
\texttt{munuSSM}. So far, the only possibility
to create an instance of the class
\texttt{BenchmarkPoint} is via the subclass
\texttt{BenchmarkPointFromFile}.
A benchmark point is initialized by doing:
\begin{lstlisting}
from munuSSM.benchmarkPointFromFile import \ BenchmarkPointFromFile
pt = BenchmarkPointFromFile(file=FILENAME)
\end{lstlisting}
Here, \texttt{FILENAME} is the path to the input file
containing the values of the free parameters.
The format of the input file is depicted in
\refli{inputfile} in \ref{inputap}.
Example input files can also be found in the
folder \texttt{example}. In the input files it
is important that the order of the lines remains
unchanged and that the parameter values start
with the first character of each line. Every character
beyond the \# sign is treated as a comment.
As already explained in \refse{secmunu},
the \textsc{Susy} parameters are \DRbar\ parameters
assumed to be given at the \texttt{Susy}-breaking
scale $M_S$, which is by default set to $1\tev$ in
the module \texttt{constants}.

Once the benchmark point \texttt{pt} is
initialized, the methods defined in
\refta{bpmeth} can be called. For example,
the tree-level spectrum and the complete set
of couplings can be obtained with:
\begin{lstlisting}
pt.calc_tree_level_spectrum()
pt.calc_tree_level_couplings()
\end{lstlisting}
Strictly speaking, only the second line
would have been sufficient, since the couplings
need the mixing matrices as input, which are
calculated when calling \texttt{calc\_tree\_level\_spectrum}.
Therefore, this function is called automatically
when \texttt{calc\_tree\_level\_couplings} is called
in case the mixing matrices are not yet available.
We can obtain the loop-corrected scalar masses
by typing:
\begin{lstlisting}
pt.calc_loop_masses(
    even=2,
    odd=1,
    momentum_mode=1)
\end{lstlisting}
Here, we explicitly set the loop order for the neutral
CP-even scalars to 2 and for the CP-odd scalars to 1.
The loop order \texttt{even=2} includes also the
contributions from the resummation of logarithmic terms
(see \refse{radcorr}).
In addition, we choose to take into account the
momentum dependence of the radiative corrections
by setting \texttt{momentum\_mode=1}, which is
the recommended value.
The values of the arguments shown above correspond
to the default values of the arguments, such that
in this case it would have been sufficient to call:
\begin{lstlisting}
pt.calc_loop_masses()
\end{lstlisting}
The loop corrected scalar masses are saved in the
class attributes \texttt{pt.Masshh\_2L} and
\texttt{pt.MassAh\_L}. The latter also contains
the mass of the unphysical Goldstone boson
with a mass of $\sim M_Z$.

The branching ratios of the neutral and charged
Higgs bosons can be obtained by calling:
\begin{lstlisting}
pt.calc_branching_ratios()
\end{lstlisting}
This will save the various branching ratios of
the neutral CP-even and CP-odd scalars
and the charged scalars in the objects:
\begin{lstlisting}
pt.BranchingRatiosh
pt.BranchingRatiosA
pt.BranchingRatiosX
\end{lstlisting}
The corresponding decay widths and also
the total decay widths are stored in the objects:
\begin{lstlisting}
pt.Gammash
pt.GammasA
pt.GammasX
\end{lstlisting}
These objects are lists of dictionaries for each
scalar particle. In the dictionaries, the different final states
are labeled by the keys, and the value corresponding
to each key is a \texttt{NumPy} array in which each index
corresponds to a family index of the final state
particles (see \refta{bpbrsatt} in \ref{return} for the
definition of each entry). For instance, the branching ratio for
the decay $h_8 \rightarrow h_1 \ h_2$ is saved in
\texttt{pt.BranchingRatiosh[7]['hhh'][0,1]}.\footnote{Indices
in python start with 0, so that the index 7
selects the particle $h_8$ etc.}
It is important to note that the same decay with the
family indices in the final state switched,
i.e., $h_8 \rightarrow h_2 \ h_1$,
is saved separately in
\texttt{pt.BranchingRatiosh[7]['hhh'][1,0]}, so that the
full branching ratio for the decay into this final state
is given by the sum. The reason for this definition is
that this allows to calculate the total decay widths
by simply summing over all elements of each array contained
in the dictionary corresponding to each particle.
The program calculates the branching ratios using the
neutral scalar masses and mixing matrices at the highest
loop level available. It will warn the user
during the calculation if only the tree-level spectrum
is used. To avoid these warnings, the user should call
\texttt{calc\_loop\_masses} before calling
\texttt{calc\_branching\_ratios}.

Finally, the collider constraints can be checked by calling
\texttt{HiggsBounds} and \texttt{HiggsSignals}:
\begin{lstlisting}
from munuSSM.higgsBounds.util import \
    check_higgsbounds_higgssignals
check_higgsbounds_higgssignals(pt)
\end{lstlisting}
One restriction is that the \texttt{HiggsBounds} libraries can
only be called once within a python session. If one wants to
check several benchmark points in the same python session,
one has to initialize them first,
save them in a list, and call the function with this list as
argument:
\begin{lstlisting}
pt_1 = BenchmarkPointFromFile(file=FILENAME_1)
pt_2 = BenchmarkPointFromFile(file=FILENAME_2)
...
pt_N = BenchmarkPointFromFile(file=FILENAME_N)
...
pts = [pt1, pt2, ..., ptN]
check_higgsbounds_higgssignals(pts)
\end{lstlisting}
To only obtain the \texttt{HiggsBounds} result, one
can call:
\begin{lstlisting}{leftmargin=.25in}
from munuSSM.higgsBounds.util import \
    check_higgsbounds
check_higgsbounds(pts)
\end{lstlisting}
An example script can be found in the file
\texttt{example.py} in the folder \texttt{example}.

\section{Numerical results: An example study for
intermediate $\tan\beta$}
\label{secexample}
To demonstrate the analysis of the Higgs sector
of the $\mu\nu$SSM using
our code, we show the results of a small parameter scan.
The parameter values correspond to the ones
given in the example input file depicted
in \ref{inputap}, except for the value of $\tan\beta$
and the values of the portal couplings $\lambda_i \equiv \lambda$.
We varied these two parameters in the range
$\tan\beta = 5\dots 20$ and $\lambda=0.02\dots 0.12$.
They are particularly relevant for the phenomenology
of the Higgs sector and the SM-like Higgs-boson mass $m_{h_1}$.

For large values of $\tan\beta$,
the radiative corrections to $m_{h_1}$ stemming
from the (s)top sector become larger, making it easier to
accommodate a mass of $m_{h_1} \sim 125\gev$. On the other hand,
the couplings of the heavy MSSM-like Higgs boson (in this scenario
$h_8$) to down-type fermions scale roughly with $\tan\beta$.
Therefore, the $b \bar b$ associated LHC cross sections are
enhanced for large values of $\tan\beta$, such that the
heavy Higgs boson cannot be too light.

The value of $\lambda$ also impacts the results in two ways.
Firstly, larger values of $\lambda$ increase the singlet-component
of the SM-like Higgs boson $h_1$. For the range of
$\tan\beta$ investigated here, this yields
a reduction of the Higgs-boson mass prediction and possibly
modifies the couplings of $h_1$ to the SM fermions and gauge bosons.
Secondly, the $\mu$-term of the MSSM is related to $\lambda$ in
the $\mu\nu$SSM. For fixed values of the singlet vevs $v_{iR}\equiv v_R =1\tev$,
we find $\mu$-values in the range
$\mu = 3 v_R \lambda / \sqrt{2} \sim 42\dots 254\gev$. The mass
of one neutral fermion (usually called Higgsino) is roughly given
by $\mu$, such that for low values of $\lambda$ the decay
$h_1 \rightarrow \chi^0_4 \chi^0_4$, with $\chi^0_4$ being the
Higgsino in this scan, becomes relevant. Furthermore, the masses
of the heavy doublet-like scalars roughly scale with
$\mu/\sin 2\beta$, such that these masses will vary over a substantial
range in this scan.

To analyze the parameter region described above, we
created input files for each benchmark point with $\tan\beta$
varying in steps of $0.2$ and $\lambda$ in steps of $0.002$.
Then, the benchmark points were initialized by creating an
instance of \texttt{BenchmarkPointFromFile} for each point.
Afterwards, we called \texttt{calc\_loop\_masses()} to obtain
the radiatively corrected neutral scalar spectrum. We used
the default options, such that the CP-even scalar masses were
calculated including the full set of higher-order corrections.
Apart from that, the default settings include the
momentum-dependence of the fixed-order
corrections at one-loop level.
Finally, we called \texttt{check\_higgsbounds\_higgssignals()}
to confront the parameter points with the current experimental
constraints. The functions mentioned above save all relevant
observables and further information in class attributes of
the instances of \texttt{BenchmarkPointFromFile}
(see \refse{secmunu}).
This information can then be easily saved to data files
using python packages like \texttt{pandas}~\cite{mckinney-proc-scipy-2010}
or graphically represented using, for instance,
the package \texttt{matplotlib}~\cite{Hunter:2007}.

\begin{figure}
\centering
\includegraphics[width=0.7\textwidth]{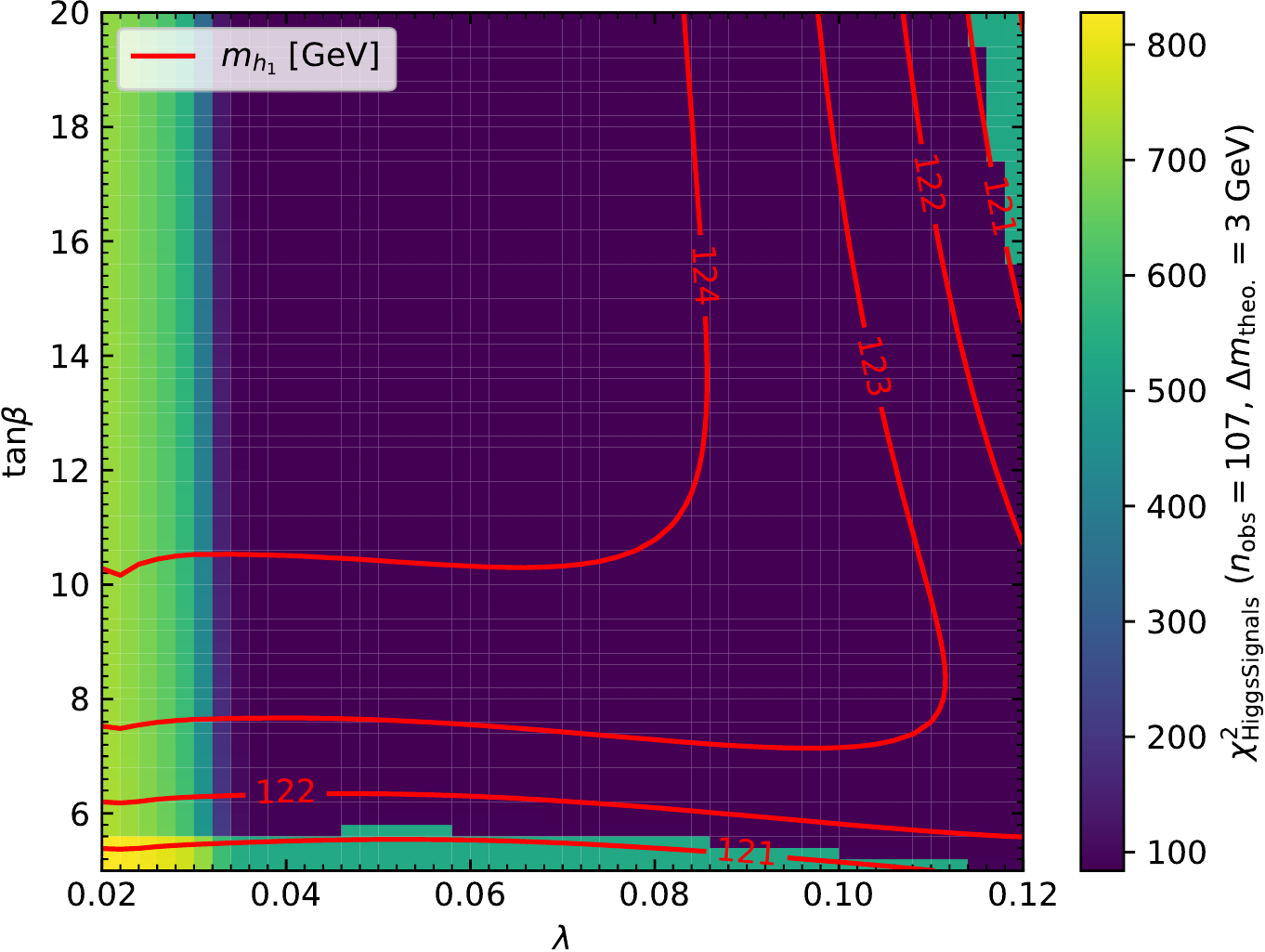}
\caption{Result of the $\chi^2$-test regarding the signal rates
of the SM Higgs boson using \texttt{HiggsSignals}.
The colour coding indicates the value of $\chi^2$.
The reference
value of the SM regarding the same set of observables
is $\chi^2_{\rm SM}=84$. The red lines are contour lines indicating
the value of the SM-like Higgs-boson mass $m_{h_1}$.}
\label{fighs}
\end{figure}

In \reffi{fighs} we summary the results related to
the SM-like Higgs boson $h_1$. We indicate the
mass $m_{h_1}$ with the red contours. Assuming a
theoretical uncertainty of $\sim 3\gev$, one can see
that a large fraction of the parameter space accommodates
the Higgs-boson mass accurately.
Only for values of $\tan\beta < 6$ and $\lambda \sim 0.12$
we find points for which $m_{h_1}$ drops below $122\gev$.
The corrections to $m_{h_1}$
stemming from contributions beyond one-loop level
are roughly of the size of $\sim 10\gev$ in this scan.
This demonstrates the importances of supplementing these
corrections via the link to \texttt{FeynHiggs}.
We can also see in \reffi{fighs} that the \texttt{HiggsSignals}
test returns low values of $\chi^2 < 100$ for $n_{\rm obs}=107$
observables in the region where
$m_{h_1}$ lies in the range $125\pm 3\gev$.
The reference value assuming the SM prediction is
$\chi^2_{\rm SM}=84$, which roughly coincides
with the values of $\chi^2$ obtained for the
$\mu\nu$SSM in the parameter space coloured in blue.
The large values of $\chi^2$ for $\lambda < 0.03$ are
caused by the decay $h_1 \rightarrow \chi^0_4 \chi^0_4$,
which becomes kinematically allowed there, spoiling the
SM-like behaviour of $h_1$.

\begin{figure}
\centering
\includegraphics[width=0.7\textwidth]{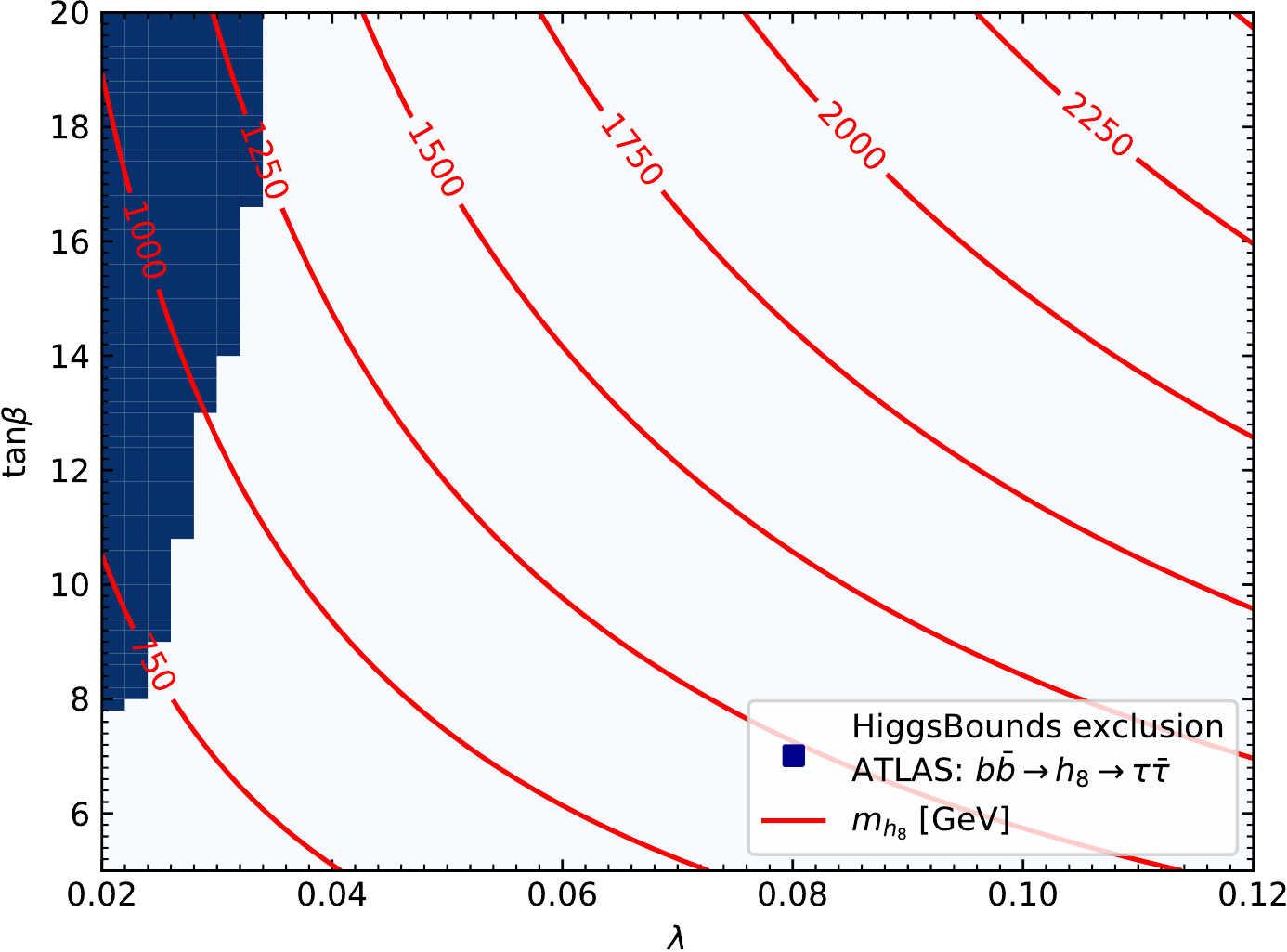}
\caption{Result of the test against collider searches for
additional Higgs bosons using \texttt{HiggsBounds}.
The blue region is excluded at $95\%$ CL due to the
search for additional Higgs bosons decaying into a
pair of $\tau$ leptons of ATLAS~\cite{Aad:2020zxo}.
The red lines are contour lines indicating the
value of the heavy Higgs-boson mass $m_{h_8}$.}
\label{fighb}
\end{figure}

In \reffi{fighb} we depict the results of the
\texttt{HiggsBounds} analysis.
In this scenario, we find one experimental search
that excludes a fraction of the parameter space
for low $\lambda$ at the 95\% confidence level.
This is related to the fact that, as explained
before, small values of $\lambda$ yield a small
effective $\mu$-parameter. This reduces also the
mass of the MSSM-like heavy Higgs boson $h_8$
(usually denoted $H$ in the MSSM).
The mass $m_{h_8}$ is indicated by the red contours
in \reffi{fighb}. The relevant experimental search
is the one for heavy Higgs bosons decaying into a
pair of $\tau$ leptons using the LHC Run~II data,
corresponding to an integrated luminosity of
$139\ \text{fb}^{-1}$, performed by
ATLAS~\cite{Aad:2020zxo}.
Both the $b \bar b$ associated production cross section
as well as the decay width of $h_8$ into a pair
of $\tau$ leptons roughly scale with $\tan^2\beta$.
Thus, larger values of $\tan\beta$
exclude parameter points even when $m_{h_8}$ is
considerably larger than $\sim 1 \tev$.
For larger values of $\lambda$,
the most sensitive experimental searches, as selected
by \texttt{HiggsBounds}, are measurements regarding
the cross section limits of the Higgs boson $h_1$
at $\sim 125\gev$. This indicates that the
predicted signal rates of the other
Higgs bosons of the $\mu\nu$SSM are substantially below
the experimental limits in most parts of the white region
of \reffi{fighb}.

\section{Conclusion and outlook}
\label{conclu}
In this paper we present the public code \texttt{munuSSM}:
A flexible python package for the phenomenological analysis
of the $\mu$-from-$\nu$ Supersymmetric Standard Model.
The code incorporates a calculation of the radiatively
corrected Higgs-boson masses. The precision of the prediction
for the SM-like Higgs-boson mass is at a comparable level
to the ones of spectrum generators for the MSSM.
This is achieved by a full one-loop renormalization
of the Higgs potential and consistently supplementing
higher-order corrections known from the MSSM via
an interface to the public code \texttt{FeynHiggs}.
For obvious reasons, this approach does not capture
effects beyond one-loop level genuine to the
$\mu\nu$SSM. For the SM-like Higgs-boson mass, these
contributions are expected to be substantially smaller
than the MSSM-like contributions considered here
for phenomenologically viable points.
Nevertheless, for an estimate of the theory
uncertainty this fact should be kept in mind.

In addition, the package \texttt{munuSSM} provides
a calculation of effective couplings and branching
ratios of the abundant scalars, pseudoscalars and
sleptons of the model. Based on these quantities,
a set of benchmark points can easily be checked
against collider constraints from the Tevatron, LEP
and the LHC via a user-friendly interface to
the public code \texttt{HiggsBounds}.
Furthermore, the presence in the spectrum of
a Higgs boson reproducing the measured signal
rates of the SM Higgs boson at $\sim 125\gev$ can be verified
via an interface to the public code
\texttt{HiggsSignals}. Since both codes are accessed
via their Fortran libraries, they can be utilized
to extract other useful quantities which would not be
directly accessible via the simpler command-line
or SLHA-file input methods. For instance, we obtain
the LHC cross sections for the neutral scalars as they
are derived within \texttt{HiggsBounds} from the
effective couplings. For a better interpretation of
the \texttt{HiggsSignals} results, we provide a SM
reference $\chi^2$ that can be taken into account when
deciding whether a benchmark point is excluded or not.

The package \texttt{munuSSM} 
is a suitable framework for the implementation
of further calculations and predictions related
to the $\mu\nu$SSM. The modular structure of the
code permits its extension without
having to know the details of the already available
features. In many cases, basic ingredients for
the implementation of new features, such as the
couplings and the mixing matrices, are already available,
providing a starting point for the exploration of
other sectors of the model (see also
Tabs.~\ref{bpinitatt}--\ref{bpbrsatt}
for more details on the model definitions).

\section*{Acknowledgements}
I thank S. Heinemeyer and C. Mu\~noz for the
collaboration in calculating the radiative
corrections and for carefully reading the
manuscript. I thank S. Brass and I. Sobolev for
testing the pre-release and providing important
feedback.
In addition, I thank H. Bahl, F. Domingo, S. Heinemeyer,
S. Pa{\ss}ehr, I. Sobolov and
G. Weiglein for helpful
correspondence regarding \texttt{FeynHiggs}.
I thank T. Stefaniak and J. Wittbrodt
for helpful correspondence regarding
\texttt{HiggsBounds} and \texttt{HiggsSignals}.
This work is supported by the Deutsche
Forschungsgemeinschaft under Germany’s Excellence
Strategy EXC2121 ``Quantum Universe'' - 390833306.

\bibliographystyle{kp}
\bibliography{lit}

\newcommand{\noop}[1]{}
\begingroup\raggedright\begin{thebibliography}{63}
\expandafter\ifx\csname natexlab\endcsname\relax\def\natexlab#1{#1}\fi

\bibitem[Lopez-Fogliani and Muñoz(2006)]{Bratchikov:2005vp}
D.~Lopez-Fogliani and C.~Muñoz, ``{Proposal for a Supersymmetric Standard
  Model}'', {\em Phys. Rev. Lett.} {\bfseries 97} (2006) 041801,
  \href{https://arxiv.org/abs/hep-ph/0508297}{{\ttfamily hep-ph/0508297}}.

\bibitem[Muñoz(2010)]{Munoz:2009an}
C.~Muñoz, ``{Phenomenology of a New Supersymmetric Standard Model: The mu nu
  SSM}'', {\em AIP Conf. Proc.} {\bfseries 1200} (2010), no.~1, 413--416,
  \href{https://arxiv.org/abs/0909.5140}{{\ttfamily arXiv:0909.5140}}.

\bibitem[Biekötter et~al.(2018)Biekötter, Heinemeyer, and
  Muñoz]{Biekotter:2017xmf}
T.~Biekötter, S.~Heinemeyer, and C.~Muñoz, ``{Precise prediction for the
  Higgs-boson masses in the $\mu\nu$SSM}'', {\em Eur. Phys. J. C} {\bfseries
  78} (2018), no.~6, 504,  \href{https://arxiv.org/abs/1712.07475}{{\ttfamily
  arXiv:1712.07475}}.

\bibitem[Biekötter et~al.(2019)Biekötter, Heinemeyer, and
  Muñoz]{Biekotter:2019gtq}
T.~Biekötter, S.~Heinemeyer, and C.~Muñoz, ``{Precise prediction for the
  Higgs-Boson masses in the $\mu\nu$SSM with three right-handed neutrino
  superfields}'', {\em Eur. Phys. J. C} {\bfseries 79} (2019), no.~8, 667,
  \href{https://arxiv.org/abs/1906.06173}{{\ttfamily arXiv:1906.06173}}.

\bibitem[Zhang(1999)]{Zhang:1998bm}
R.-J. Zhang, ``{Two loop effective potential calculation of the lightest CP
  even Higgs boson mass in the MSSM}'', {\em Phys. Lett. B} {\bfseries 447}
  (1999) 89--97,  \href{https://arxiv.org/abs/hep-ph/9808299}{{\ttfamily
  hep-ph/9808299}}.

\bibitem[Espinosa and Zhang(2000)]{Espinosa:1999zm}
J.~R. Espinosa and R.-J. Zhang, ``{MSSM lightest CP even Higgs boson mass to
  O(alpha(s) alpha(t)): The Effective potential approach}'', {\em JHEP}
  {\bfseries 03} (2000) 026,
  \href{https://arxiv.org/abs/hep-ph/9912236}{{\ttfamily hep-ph/9912236}}.

\bibitem[Heinemeyer et~al.(1999)Heinemeyer, Hollik, and
  Weiglein]{Heinemeyer:1998np}
S.~Heinemeyer, W.~Hollik, and G.~Weiglein, ``{The Masses of the neutral CP -
  even Higgs bosons in the MSSM: Accurate analysis at the two loop level}'',
  {\em Eur. Phys. J. C} {\bfseries 9} (1999) 343--366,
  \href{https://arxiv.org/abs/hep-ph/9812472}{{\ttfamily hep-ph/9812472}}.

\bibitem[Kpatcha et~al.(2020)Kpatcha, Ruiz~de Austri, López-Fogliani, and
  Muñoz]{Kpatcha:2019qsz}
E.~Kpatcha, R.~Ruiz~de Austri, D.~E. López-Fogliani, and C.~Muñoz, ``{Impact
  of Higgs physics on the parameter space of the $\mu \nu \mathrm{SSM}$}'',
  {\em Eur. Phys. J. C} {\bfseries 80} (2020), no.~4, 336,
  \href{https://arxiv.org/abs/1910.08062}{{\ttfamily arXiv:1910.08062}}.

\bibitem[Ghosh et~al.(2018)Ghosh, Lara, Lopez-Fogliani, Muñoz, and Ruiz~de
  Austri]{Ghosh:2017yeh}
P.~Ghosh, I.~Lara, D.~E. Lopez-Fogliani, C.~Muñoz, and R.~Ruiz~de Austri,
  ``{Searching for left sneutrino LSP at the LHC}'', {\em Int. J. Mod. Phys. A}
  {\bfseries 33} (2018), no.~18n19, 1850110,
  \href{https://arxiv.org/abs/1707.02471}{{\ttfamily arXiv:1707.02471}}.

\bibitem[Lara et~al.(2018)Lara, López-Fogliani, Muñoz, Nagata, Otono, and
  Ruiz De~Austri]{Lara:2018rwv}
I.~Lara, D.~E. López-Fogliani, C.~Muñoz, N.~Nagata, H.~Otono, and R.~Ruiz
  De~Austri, ``{Looking for the left sneutrino LSP with displaced-vertex
  searches}'', {\em Phys. Rev. D} {\bfseries 98} (2018), no.~7, 075004,
  \href{https://arxiv.org/abs/1804.00067}{{\ttfamily arXiv:1804.00067}}.

\bibitem[Kpatcha et~al.(2019)Kpatcha, Lara, López-Fogliani, Muñoz, Nagata,
  Otono, and Ruiz De~Austri]{Kpatcha:2019gmq}
E.~Kpatcha, I.~Lara, D.~E. López-Fogliani, C.~Muñoz, N.~Nagata, H.~Otono, and
  R.~Ruiz De~Austri, ``{Sampling the $\mu \nu $SSM for displaced decays of the
  tau left sneutrino LSP at the LHC}'', {\em Eur. Phys. J. C} {\bfseries 79}
  (2019), no.~11, 934,  \href{https://arxiv.org/abs/1907.02092}{{\ttfamily
  arXiv:1907.02092}}.

\bibitem[Porod(2003)]{Porod:2003um}
W.~Porod, ``{SPheno, a program for calculating supersymmetric spectra, SUSY
  particle decays and SUSY particle production at e+ e- colliders}'', {\em
  Comput. Phys. Commun.} {\bfseries 153} (2003) 275--315,
  \href{https://arxiv.org/abs/hep-ph/0301101}{{\ttfamily hep-ph/0301101}}.

\bibitem[Porod and Staub(2012)]{Porod:2011nf}
W.~Porod and F.~Staub, ``{SPheno 3.1: Extensions including flavour, CP-phases
  and models beyond the MSSM}'', {\em Comput. Phys. Commun.} {\bfseries 183}
  (2012) 2458--2469,  \href{https://arxiv.org/abs/1104.1573}{{\ttfamily
  arXiv:1104.1573}}.

\bibitem[Staub(2010)]{Staub:2009bi}
F.~Staub, ``{From Superpotential to Model Files for FeynArts and
  CalcHep/CompHep}'', {\em Comput. Phys. Commun.} {\bfseries 181} (2010)
  1077--1086,  \href{https://arxiv.org/abs/0909.2863}{{\ttfamily
  arXiv:0909.2863}}.

\bibitem[Staub(2011)]{Staub:2010jh}
F.~Staub, ``{Automatic Calculation of supersymmetric Renormalization Group
  Equations and Self Energies}'', {\em Comput. Phys. Commun.} {\bfseries 182}
  (2011) 808--833,  \href{https://arxiv.org/abs/1002.0840}{{\ttfamily
  arXiv:1002.0840}}.

\bibitem[Staub(2013)]{Staub:2012pb}
F.~Staub, ``{SARAH 3.2: Dirac Gauginos, UFO output, and more}'', {\em Comput.
  Phys. Commun.} {\bfseries 184} (2013) 1792--1809,
  \href{https://arxiv.org/abs/1207.0906}{{\ttfamily arXiv:1207.0906}}.

\bibitem[Staub(2014)]{Staub:2013tta}
F.~Staub, ``{SARAH 4 : A tool for (not only SUSY) model builders}'', {\em
  Comput. Phys. Commun.} {\bfseries 185} (2014) 1773--1790,
 \href{https://arxiv.org/abs/1309.7223}{{\ttfamily arXiv:1309.7223}}.

\bibitem[Athron et~al.(2015)Athron, Park, St\"ockinger, and
  Voigt]{Athron:2014yba}
P.~Athron, J.-h. Park, D.~St\"ockinger, and A.~Voigt,
  ``{FlexibleSUSY\textemdash{}A spectrum generator generator for supersymmetric
  models}'', {\em Comput. Phys. Commun.} {\bfseries 190} (2015) 139--172,
  \href{https://arxiv.org/abs/1406.2319}{{\ttfamily arXiv:1406.2319}}.

\bibitem[Athron et~al.(2017)Athron, Park, Steudtner, St\"ockinger, and
  Voigt]{Athron:2016fuq}
P.~Athron, J.-h. Park, T.~Steudtner, D.~St\"ockinger, and A.~Voigt, ``{Precise
  Higgs mass calculations in (non-)minimal supersymmetry at both high and low
  scales}'', {\em JHEP} {\bfseries 01} (2017) 079,
  \href{https://arxiv.org/abs/1609.00371}{{\ttfamily arXiv:1609.00371}}.

\bibitem[Athron et~al.(2018)Athron, Bach, Harries, Kwasnitza, Park,
  St\"ockinger, Voigt, and Ziebell]{Athron:2017fvs}
P.~Athron, M.~Bach, D.~Harries, T.~Kwasnitza, J.-h. Park, D.~St\"ockinger,
  A.~Voigt, and J.~Ziebell, ``{FlexibleSUSY 2.0: Extensions to investigate the
  phenomenology of SUSY and non-SUSY models}'', {\em Comput. Phys. Commun.}
  {\bfseries 230} (2018) 145--217,
  \href{https://arxiv.org/abs/1710.03760}{{\ttfamily arXiv:1710.03760}}.

\bibitem[Heinemeyer et~al.(2000)Heinemeyer, Hollik, and
  Weiglein]{Heinemeyer:1998yj}
S.~Heinemeyer, W.~Hollik, and G.~Weiglein, ``{FeynHiggs: A Program for the
  calculation of the masses of the neutral CP even Higgs bosons in the MSSM}'',
  {\em Comput. Phys. Commun.} {\bfseries 124} (2000) 76--89,
  \href{https://arxiv.org/abs/hep-ph/9812320}{{\ttfamily hep-ph/9812320}}.

\bibitem[Degrassi et~al.(2003)Degrassi, Heinemeyer, Hollik, Slavich, and
  Weiglein]{Degrassi:2002fi}
G.~Degrassi, S.~Heinemeyer, W.~Hollik, P.~Slavich, and G.~Weiglein, ``{Towards
  high precision predictions for the MSSM Higgs sector}'', {\em Eur. Phys. J.
  C} {\bfseries 28} (2003) 133--143,
  \href{https://arxiv.org/abs/hep-ph/0212020}{{\ttfamily hep-ph/0212020}}.

\bibitem[Frank et~al.(2007)Frank, Hahn, Heinemeyer, Hollik, Rzehak, and
  Weiglein]{Frank:2006yh}
M.~Frank, T.~Hahn, S.~Heinemeyer, W.~Hollik, H.~Rzehak, and G.~Weiglein, ``{The
  Higgs Boson Masses and Mixings of the Complex MSSM in the
  Feynman-Diagrammatic Approach}'', {\em JHEP} {\bfseries 02} (2007) 047,
  \href{https://arxiv.org/abs/hep-ph/0611326}{{\ttfamily hep-ph/0611326}}.

\bibitem[Hahn et~al.(2014)Hahn, Heinemeyer, Hollik, Rzehak, and
  Weiglein]{Hahn:2013ria}
T.~Hahn, S.~Heinemeyer, W.~Hollik, H.~Rzehak, and G.~Weiglein,
  ``{High-Precision Predictions for the Light CP -Even Higgs Boson Mass of the
  Minimal Supersymmetric Standard Model}'', {\em Phys. Rev. Lett.} {\bfseries
  112} (2014), no.~14, 141801,
  \href{https://arxiv.org/abs/1312.4937}{{\ttfamily arXiv:1312.4937}}.

\bibitem[Bahl and Hollik(2016)]{Bahl:2016brp}
H.~Bahl and W.~Hollik, ``{Precise prediction for the light MSSM Higgs boson
  mass combining effective field theory and fixed-order calculations}'', {\em
  Eur. Phys. J. C} {\bfseries 76} (2016), no.~9, 499,
  \href{https://arxiv.org/abs/1608.01880}{{\ttfamily arXiv:1608.01880}}.

\bibitem[Bahl et~al.(2018)Bahl, Heinemeyer, Hollik, and Weiglein]{Bahl:2017aev}
H.~Bahl, S.~Heinemeyer, W.~Hollik, and G.~Weiglein, ``{Reconciling EFT and
  hybrid calculations of the light MSSM Higgs-boson mass}'', {\em Eur. Phys. J.
  C} {\bfseries 78} (2018), no.~1, 57,
  \href{https://arxiv.org/abs/1706.00346}{{\ttfamily arXiv:1706.00346}}.

\bibitem[Bahl et~al.(2020)Bahl, Hahn, Heinemeyer, Hollik, Paßehr, Rzehak, and
  Weiglein]{Bahl:2018qog}
H.~Bahl, T.~Hahn, S.~Heinemeyer, W.~Hollik, S.~Paßehr, H.~Rzehak, and
  G.~Weiglein, ``{Precision calculations in the MSSM Higgs-boson sector with
  FeynHiggs 2.14}'', {\em Comput. Phys. Commun.} {\bfseries 249} (2020) 107099,
   \href{https://arxiv.org/abs/1811.09073}{{\ttfamily arXiv:1811.09073}}.

\bibitem[Heinemeyer et~al.(2013)Heinemeyer, Mariotti, Passarino, Tanaka,
  et~al.]{Heinemeyer:2013tqa}
{\bfseries LHC Higgs Cross Section Working Group} Collaboration, S.~Heinemeyer,
  C.~Mariotti, G.~Passarino, R.~Tanaka, {\em et~al.}, ``{Handbook of LHC Higgs
  Cross Sections: 3. Higgs Properties}'',
  \href{https://arxiv.org/abs/1307.1347}{{\ttfamily arXiv:1307.1347}}.

\bibitem[de~Florian et~al.(2016)]{deFlorian:2016spz}
{\bfseries LHC Higgs Cross Section Working Group} Collaboration, D.~de~Florian
  {\em et~al.}, ``{Handbook of LHC Higgs Cross Sections: 4. Deciphering the
  Nature of the Higgs Sector}'',
  \href{https://arxiv.org/abs/1610.07922}{{\ttfamily arXiv:1610.07922}}.

\bibitem[Bechtle et~al.(2010)Bechtle, Brein, Heinemeyer, Weiglein, and
  Williams]{Bechtle:2008jh}
P.~Bechtle, O.~Brein, S.~Heinemeyer, G.~Weiglein, and K.~E. Williams,
  ``{HiggsBounds: Confronting Arbitrary Higgs Sectors with Exclusion Bounds
  from LEP and the Tevatron}'', {\em Comput. Phys. Commun.} {\bfseries 181}
  (2010) 138--167,  \href{https://arxiv.org/abs/0811.4169}{{\ttfamily
  arXiv:0811.4169}}.

\bibitem[Bechtle et~al.(2011)Bechtle, Brein, Heinemeyer, Weiglein, and
  Williams]{Bechtle:2011sb}
P.~Bechtle, O.~Brein, S.~Heinemeyer, G.~Weiglein, and K.~E. Williams,
  ``{HiggsBounds 2.0.0: Confronting Neutral and Charged Higgs Sector
  Predictions with Exclusion Bounds from LEP and the Tevatron}'', {\em Comput.
  Phys. Commun.} {\bfseries 182} (2011) 2605--2631,
  \href{https://arxiv.org/abs/1102.1898}{{\ttfamily arXiv:1102.1898}}.

\bibitem[Bechtle et~al.(2012)Bechtle, Brein, Heinemeyer, Stal, Stefaniak,
  Weiglein, and Williams]{Bechtle:2013gu}
P.~Bechtle, O.~Brein, S.~Heinemeyer, O.~Stal, T.~Stefaniak, G.~Weiglein, and
  K.~Williams, ``{Recent Developments in HiggsBounds and a Preview of
  HiggsSignals}'', {\em PoS} {\bfseries CHARGED2012} (2012) 024,
  \href{https://arxiv.org/abs/1301.2345}{{\ttfamily arXiv:1301.2345}}.

\bibitem[Bechtle et~al.(2014)Bechtle, Brein, Heinemeyer, Stål, Stefaniak,
  Weiglein, and Williams]{Bechtle:2013wla}
P.~Bechtle, O.~Brein, S.~Heinemeyer, O.~Stål, T.~Stefaniak, G.~Weiglein, and
  K.~E. Williams, ``{$\mathsf{HiggsBounds}-4$: Improved Tests of Extended Higgs
  Sectors against Exclusion Bounds from LEP, the Tevatron and the LHC}'', {\em
  Eur. Phys. J. C} {\bfseries 74} (2014), no.~3, 2693,
  \href{https://arxiv.org/abs/1311.0055}{{\ttfamily arXiv:1311.0055}}.

\bibitem[Bechtle et~al.(2015)Bechtle, Heinemeyer, Stal, Stefaniak, and
  Weiglein]{Bechtle:2015pma}
P.~Bechtle, S.~Heinemeyer, O.~Stal, T.~Stefaniak, and G.~Weiglein, ``{Applying
  Exclusion Likelihoods from LHC Searches to Extended Higgs Sectors}'', {\em
  Eur. Phys. J. C} {\bfseries 75} (2015), no.~9, 421,
  \href{https://arxiv.org/abs/1507.06706}{{\ttfamily arXiv:1507.06706}}.

\bibitem[Bechtle et~al.(2020)Bechtle, Dercks, Heinemeyer, Klingl, Stefaniak,
  Weiglein, and Wittbrodt]{Bechtle:2020pkv}
P.~Bechtle, D.~Dercks, S.~Heinemeyer, T.~Klingl, T.~Stefaniak, G.~Weiglein, and
  J.~Wittbrodt, ``{HiggsBounds-5: Testing Higgs Sectors in the LHC 13 TeV
  Era}'',  \href{https://arxiv.org/abs/2006.06007}{{\ttfamily
  arXiv:2006.06007}}.

\bibitem[Bechtle et~al.(2014)Bechtle, Heinemeyer, Stål, Stefaniak, and
  Weiglein]{Bechtle:2013xfa}
P.~Bechtle, S.~Heinemeyer, O.~Stål, T.~Stefaniak, and G.~Weiglein,
  ``{$HiggsSignals$: Confronting arbitrary Higgs sectors with measurements at
  the Tevatron and the LHC}'', {\em Eur. Phys. J. C} {\bfseries 74} (2014),
  no.~2, 2711,  \href{https://arxiv.org/abs/1305.1933}{{\ttfamily
  arXiv:1305.1933}}.

\bibitem[Stål and Stefaniak(2013)]{Stal:2013hwa}
O.~Stål and T.~Stefaniak, ``{Constraining extended Higgs sectors with
  HiggsSignals}'', {\em PoS} {\bfseries EPS-HEP2013} (2013) 314,
  \href{https://arxiv.org/abs/1310.4039}{{\ttfamily arXiv:1310.4039}}.

\bibitem[Bechtle et~al.(2014)Bechtle, Heinemeyer, Stål, Stefaniak, and
  Weiglein]{Bechtle:2014ewa}
P.~Bechtle, S.~Heinemeyer, O.~Stål, T.~Stefaniak, and G.~Weiglein, ``{Probing
  the Standard Model with Higgs signal rates from the Tevatron, the LHC and a
  future ILC}'', {\em JHEP} {\bfseries 11} (2014) 039,
  \href{https://arxiv.org/abs/1403.1582}{{\ttfamily arXiv:1403.1582}}.

\bibitem[Bechtle et~al.(2020)Bechtle, Heinemeyer, Klingl, Stefaniak, Weiglein,
  and Wittbrodt]{Bechtle:2020uwn}
P.~Bechtle, S.~Heinemeyer, T.~Klingl, T.~Stefaniak, G.~Weiglein, and
  J.~Wittbrodt, ``{HiggsSignals-2: Probing new physics with precision Higgs
  measurements in the LHC 13 TeV era}'',
  \href{https://arxiv.org/abs/2012.09197}{{\ttfamily arXiv:2012.09197}}.

\bibitem[Lopez-Fogliani and Mu\~noz(2020)]{Lopez-Fogliani:2020gzo}
D.~E. Lopez-Fogliani and C.~Mu\~noz, ``{Searching for Supersymmetry: The
  $\mu\nu$SSM}'',  \href{https://arxiv.org/abs/2009.01380}{{\ttfamily
  arXiv:2009.01380}}.

\bibitem[Brignole et~al.(1998)Brignole, Ibanez, and Muñoz]{Brignole:1997dp}
A.~Brignole, L.~E. Ibanez, and C.~Muñoz, ``{Soft supersymmetry breaking terms
  from supergravity and superstring models}'', {\em Adv. Ser. Direct. High
  Energy Phys.} {\bfseries 18} (1998) 125--148,
 \href{https://arxiv.org/abs/hep-ph/9707209}{{\ttfamily arXiv:hep-ph/9707209}}.

\bibitem[Biekötter(2019)]{Biekotter:2019gtp}
T.~Biekötter, ``{Phenomenology of the Higgs sectors of the $\mu\nu$SSM and the
  N2HDM}'', PhD thesis, U. Autonoma, Madrid (main), 2019.

\bibitem[Hahn(2001)]{Hahn:2000kx}
T.~Hahn, ``{Generating Feynman diagrams and amplitudes with FeynArts 3}'', {\em
  Comput. Phys. Commun.} {\bfseries 140} (2001) 418--431,
 \href{https://arxiv.org/abs/hep-ph/0012260}{{\ttfamily arXiv:hep-ph/0012260}}.

\bibitem[Gómez-Vargas et~al.(2020)Gómez-Vargas, López-Fogliani, Muñoz, and
  Perez]{Gomez-Vargas:2019vci}
G.~A. Gómez-Vargas, D.~E. López-Fogliani, C.~Muñoz, and A.~D. Perez,
  ``{MeV-GeV $\gamma$-ray telescopes probing axino LSP/gravitino NLSP as dark
  matter in the $\mu\nu$SSM}'', {\em JCAP} {\bfseries 01} (2020) 058,
  \href{https://arxiv.org/abs/1911.03191}{{\ttfamily arXiv:1911.03191}}.

\bibitem[Gómez-Vargas et~al.(2019)Gómez-Vargas, López-Fogliani, Muñoz, and
  Perez]{Gomez-Vargas:2019mqk}
G.~A. Gómez-Vargas, D.~E. López-Fogliani, C.~Muñoz, and A.~D. Perez,
  ``{MeV-GeV $\gamma$-ray telescopes probing gravitino LSP with coexisting
  axino NLSP as dark matter in the $\mu\nu$SSM}'',
  \href{https://arxiv.org/abs/1911.08550}{{\ttfamily arXiv:1911.08550}}.

\bibitem[Hahn and Perez-Victoria(1999)]{Hahn:1998yk}
T.~Hahn and M.~Perez-Victoria, ``{Automatized one loop calculations in
  four-dimensions and D-dimensions}'', {\em Comput. Phys. Commun.} {\bfseries
  118} (1999) 153--165,  \href{https://arxiv.org/abs/hep-ph/9807565}{{\ttfamily
  hep-ph/9807565}}.

\bibitem[Draper and Rzehak(2016)]{Draper:2016pys}
P.~Draper and H.~Rzehak, ``{A Review of Higgs Mass Calculations in
  Supersymmetric Models}'', {\em Phys. Rept.} {\bfseries 619} (2016) 1--24,
 \href{https://arxiv.org/abs/1601.01890}{{\ttfamily arXiv:1601.01890}}.

\bibitem[Ellwanger et~al.(2010)Ellwanger, Hugonie, and
  Teixeira]{Ellwanger:2009dp}
U.~Ellwanger, C.~Hugonie, and A.~M. Teixeira, ``{The Next-to-Minimal
  Supersymmetric Standard Model}'', {\em Phys. Rept.} {\bfseries 496} (2010)
  1--77,  \href{https://arxiv.org/abs/0910.1785}{{\ttfamily arXiv:0910.1785}}.

\bibitem[Harris et~al.(2020)]{Harris:2020xlr}
C.~R. Harris {\em et~al.}, ``{Array programming with NumPy}'', {\em Nature}
  {\bfseries 585} (2020), no.~7825, 357--362,
  \href{https://arxiv.org/abs/2006.10256}{{\ttfamily arXiv:2006.10256}}.

\bibitem[Djouadi et~al.(1998)Djouadi, Kalinowski, and Spira]{Djouadi:1997yw}
A.~Djouadi, J.~Kalinowski, and M.~Spira, ``{HDECAY: A Program for Higgs boson
  decays in the standard model and its supersymmetric extension}'', {\em
  Comput. Phys. Commun.} {\bfseries 108} (1998) 56--74,
  \href{https://arxiv.org/abs/hep-ph/9704448}{{\ttfamily hep-ph/9704448}}.

\bibitem[Spira(1998)]{Spira:1997dg}
M.~Spira, ``{QCD effects in Higgs physics}'', {\em Fortsch. Phys.} {\bfseries
  46} (1998) 203--284,  \href{https://arxiv.org/abs/hep-ph/9705337}{{\ttfamily
  hep-ph/9705337}}.

\bibitem[Butterworth et~al.(2010)]{Butterworth:2010ym}
J.~Butterworth {\em et~al.}, ``{THE TOOLS AND MONTE CARLO WORKING GROUP Summary
  Report from the Les Houches 2009 Workshop on TeV Colliders}'', in ``{6th Les
  Houches Workshop on Physics at TeV Colliders}''.
\newblock 3 2010.
\newblock  \href{https://arxiv.org/abs/1003.1643}{{\ttfamily arXiv:1003.1643}}.

\bibitem[Bredenstein et~al.(2006)Bredenstein, Denner, Dittmaier, and
  Weber]{Bredenstein:2006rh}
A.~Bredenstein, A.~Denner, S.~Dittmaier, and M.~Weber, ``{Precise predictions
  for the Higgs-boson decay H $\rightarrow$ WW/ZZ $\rightarrow$ 4 leptons}'',
  {\em Phys. Rev. D} {\bfseries 74} (2006) 013004,
  \href{https://arxiv.org/abs/hep-ph/0604011}{{\ttfamily hep-ph/0604011}}.

\bibitem[Bredenstein et~al.(2007)Bredenstein, Denner, Dittmaier, and
  Weber]{Bredenstein:2006ha}
A.~Bredenstein, A.~Denner, S.~Dittmaier, and M.~Weber, ``{Radiative corrections
  to the semileptonic and hadronic Higgs-boson decays H $\rightarrow$ W W / Z Z
  $\rightarrow$ 4 fermions}'', {\em JHEP} {\bfseries 02} (2007) 080,
  \href{https://arxiv.org/abs/hep-ph/0611234}{{\ttfamily hep-ph/0611234}}.

\bibitem[Goodsell et~al.(2017)Goodsell, Liebler, and Staub]{Goodsell:2017pdq}
M.~D. Goodsell, S.~Liebler, and F.~Staub, ``{Generic calculation of two-body
  partial decay widths at the full one-loop level}'', {\em Eur. Phys. J. C}
  {\bfseries 77} (2017), no.~11, 758,
  \href{https://arxiv.org/abs/1703.09237}{{\ttfamily arXiv:1703.09237}}.

\bibitem[Spira(2017)]{Spira:2016ztx}
M.~Spira, ``{Higgs Boson Production and Decay at Hadron Colliders}'', {\em
  Prog. Part. Nucl. Phys.} {\bfseries 95} (2017) 98--159,
  \href{https://arxiv.org/abs/1612.07651}{{\ttfamily arXiv:1612.07651}}.

\bibitem[Drechsel et~al.(2017)Drechsel, Gr\"ober, Heinemeyer, Muhlleitner,
  Rzehak, and Weiglein]{Drechsel:2016htw}
P.~Drechsel, R.~Gr\"ober, S.~Heinemeyer, M.~M. Muhlleitner, H.~Rzehak, and
  G.~Weiglein, ``{Higgs-Boson Masses and Mixing Matrices in the NMSSM: Analysis
  of On-Shell Calculations}'', {\em Eur. Phys. J. C} {\bfseries 77} (2017),
  no.~6, 366,  \href{https://arxiv.org/abs/1612.07681}{{\ttfamily
  arXiv:1612.07681}}.

\bibitem[Hollik et~al.(2019{\natexlab{a}})Hollik, Liebler, Moortgat-Pick,
  Pa\ss{}ehr, and Weiglein]{Hollik:2018yek}
W.~G. Hollik, S.~Liebler, G.~Moortgat-Pick, S.~Pa\ss{}ehr, and G.~Weiglein,
  ``{Phenomenology of the inflation-inspired NMSSM at the electroweak scale}'',
  {\em Eur. Phys. J. C} {\bfseries 79} (2019){\natexlab{a}}, no.~1, 75,
  \href{https://arxiv.org/abs/1809.07371}{{\ttfamily arXiv:1809.07371}}.

\bibitem[Hollik et~al.(2019{\natexlab{b}})Hollik, Liebler, Moortgat-Pick,
  Pa\ss{}ehr, and Weiglein]{Hollik:2018bwj}
W.~G. Hollik, S.~Liebler, G.~Moortgat-Pick, S.~Pa\ss{}ehr, and G.~Weiglein,
  ``{Phenomenological consequences of Higgs inflation in the NMSSM at the
  electroweak scale}'', {\em PoS} {\bfseries ICHEP2018} (2019){\natexlab{b}}
  455,  \href{https://arxiv.org/abs/1811.12838}{{\ttfamily arXiv:1811.12838}}.

\bibitem[Hahn(2006)]{Hahn:2006hr}
T.~Hahn, ``{Routines for the diagonalization of complex matrices}'',
  \href{https://arxiv.org/abs/physics/0607103}{{\ttfamily physics/0607103}}.

\bibitem[{W}es {M}c{K}inney(2010)]{mckinney-proc-scipy-2010}
{W}es {M}c{K}inney, ``{D}ata {S}tructures for {S}tatistical {C}omputing in
  {P}ython'', in ``{P}roceedings of the 9th {P}ython in {S}cience
  {C}onference'', {S}t\'efan van~der {W}alt and {J}arrod {M}illman, eds.,
  pp.~56 -- 61.
\newblock 2010.

\bibitem[Hunter(2007)]{Hunter:2007}
J.~D. Hunter, ``Matplotlib: A 2d graphics environment'', {\em Computing in
  Science \& Engineering} {\bfseries 9} (2007), no.~3, 90--95.

\bibitem[Aad et~al.(2020)]{Aad:2020zxo}
{\bfseries ATLAS} Collaboration, G.~Aad {\em et~al.}, ``{Search for heavy Higgs
  bosons decaying into two tau leptons with the ATLAS detector using $pp$
  collisions at $\sqrt{s}=13$ TeV}'', {\em Phys. Rev. Lett.} {\bfseries 125}
  (2020), no.~5, 051801,  \href{https://arxiv.org/abs/2002.12223}{{\ttfamily
  arXiv:2002.12223}}.

\end{thebibliography}\endgroup

\newpage

\appendix

\section{Example input file}
\label{inputap}

\begin{lstlisting}[label=inputfile, basicstyle=\ttfamily\footnotesize,
    caption=Example input file with random parameter values, frame=tb]
# munuSSM SUSY PARAMETERS ###############################
15.0            # TanBe
0.0005          # vL_1
0.0005          # vL_2
0.0005          # vL_3
1000.0          # vR_1
1000.0          # vR_2
1000.0          # vR_3
0.08            # lam_1
0.08            # lam_2
0.08            # lam_3
0.3             # kap_111
0.0             # kap_112
0.0             # kap_113
0.0             # kap_122
0.0             # kap_123
0.0             # kap_133
0.3             # kap_222
0.0             # kap_223
0.0             # kap_233
0.3             # kap_333
1.0e-07         # Yv_11
0.0             # Yv_12
0.0             # Yv_13
0.0             # Yv_21
1.0e-07         # Yv_22
0.0             # Yv_23
0.0             # Yv_31
0.0             # Yv_32
1.0e-07         # Yv_33
# munuSSM SOFT PARAMETERS ###############################
0.0             # ml2_12
0.0             # ml2_13
0.0             # ml2_23
0.0             # mlHd2_1
0.0             # mlHd2_2
0.0             # mlHd2_3
0.0             # mv2_12
0.0             # mv2_13
0.0             # mv2_23
2250000.0       # mq2_11
2250000.0       # mq2_22
2250000.0       # mq2_33
2250000.0       # mu2_11
2250000.0       # mu2_22
2250000.0       # mu2_33
2250000.0       # md2_11
2250000.0       # md2_22
2250000.0       # md2_33
2250000.0       # me2_11
0.0             # me2_12
0.0             # me2_13
2250000.0       # me2_22
0.0             # me2_23
2250000.0       # me2_33
1000.0         # Au_11
1000.0         # Au_22
2800.0         # Au_33
1000.0         # Ad_11
1000.0         # Ad_22
1000.0         # Ad_33
1000.0         # Ae_11
0.0             # Ae_12
0.0             # Ae_13
0.0             # Ae_21
1000.0         # Ae_22
0.0             # Ae_23
0.0             # Ae_31
0.0             # Ae_32
1000.0         # Ae_33
-1000.0         # Av_11
0.0             # Av_12
0.0             # Av_13
0.0             # Av_21
-1000.0         # Av_22
0.0             # Av_23
0.0             # Av_31
0.0             # Av_32
-1000.0         # Av_33
1000.0           # Alam_1
1000.0           # Alam_2
1000.0           # Alam_3
-100.0          # Akap_111
0.0             # Akap_112
0.0             # Akap_113
0.0             # Akap_122
0.0             # Akap_123
0.0             # Akap_133
-100.0          # Akap_222
0.0             # Akap_223
0.0             # Akap_233
-100.0          # Akap_333
300.0           # M1
500.0           # M2
1700.0          # M3
\end{lstlisting}

\newpage

\section{Return values and class attributes}
\label{return}

In the following tables we list the attributes
that are set for an instance of the class
\texttt{BenchmarkPoint} and the return values
for each method defined in the class.
In most cases the objects listed in the tables
are of type \texttt{NumberQP} or \texttt{ArrayQP},
as defined in the module \texttt{dataObjects}
(see \refse{secmunu}). We remind the reader that the
values can be obtained in terms of regular
floats or
\texttt{NumPy} float arrays by typing
\texttt{a.float} if \texttt{a} is of type
\texttt{NumberQP} or \texttt{ArrayQP}.

{\renewcommand{\arraystretch}{1.2}
\footnotesize
\begin{longtable}{p{0.12\textwidth}p{0.12\textwidth}p{0.62\textwidth}}
{\normalsize \texttt{self.\_\_init\_\_(file)}} & \\
\hline
\texttt{Delta} &  &
Value of the UV divergent piece of loop integrals
$\Delta = 1/\varepsilon^{\mathrm{UV}} $ \\
\texttt{MT\_POLE} &  &
Top quark pole mass $M_t$ \\
\texttt{MB\_MB} &  &
Bottom quark mass at the scale $\overline m_b(m_b)$ \\
\texttt{ScaleFac} &  &
Renormalization scale in powers of $M_t$:
$\mu_R / M_t$ \\
\texttt{DRbarScale} &  &
Input scale of \DRbar\ parameters $\mu_0$ \\
\texttt{MW} &  &
Mass of the $W$ boson $M_W$ \\
\texttt{MZ} &  &
Mass of the $Z$ boson $M_Z$ \\
\texttt{GF} &  &
Fermi constant $G_F$ \\
\texttt{AlfaS} &  &
Strong QCD coupling constant $\alpha_S(M_Z)$ \\
\texttt{MC} &  &
Charme quark mass $m_c$ \\
\texttt{MS} &  &
Strange quark mass $m_s$ \\
\texttt{MU} &  &
Up quark mass $m_u$ \\
\texttt{MD} &  &
Down quark mass $m_d$ \\
\texttt{ML} &  &
Tauon mass $m_\tau$ \\
\texttt{MM} &  &
Muon mass $m_\mu$ \\
\texttt{ME} &  &
Electron mass $m_e$ \\
\texttt{MT} &  &
SM-\MSbar\ top quark mass $\overline{m}_t^{\MSbar,\rm SM}(M_t)$ \\
\texttt{MB} &  &
MSSM-\DRbar\ bottom quark mass $\overline{m}_b^{\DRbar,\rm MSSM}(M_S)$ \\
\texttt{TB} &  &
Ratio of doublet vevs $\tan\beta$ \\
\texttt{vL} & \texttt{(3, )} &
Left-handed sneutrino vevs $v_{iL}$ \\
\texttt{vR} & \texttt{(3, )} &
Right-handed sneutrino vevs $v_{iR}$ \\
\texttt{lam} & \texttt{(3, )} &
Superpotential couplings $\lambda_i$ \\
\texttt{kap} & \texttt{(3, 3, 3)} &
Superpotential couplings $\kappa_{ijk}$ \\
\texttt{Yv} & \texttt{(3, 3)} &
Neutrino Yukawa couplings $Y^\nu_{ij}$ \\
\texttt{mlHd2} & \texttt{(3, )} &
Soft mass parameters $(m_{H_d \widetilde{L}}^2)_i$ \\
\texttt{mq2} & \texttt{(3, 3)} &
Soft mass parameters $(m_{\widetilde{Q}}^2)_{ij}$ \\
\texttt{mu2} & \texttt{(3, 3)} &
Soft mass parameters $(m_{\widetilde{u}}^2)_{ij}$ \\
\texttt{md2} & \texttt{(3, 3)} &
Soft mass parameters $(m_{\widetilde{d}}^2)_{ij}$ \\
\texttt{me2} & \texttt{(3, 3)} &
Soft mass parameters $(m_{\widetilde{e}}^2)_{ij}$ \\
\texttt{Au} & \texttt{(3, 3)} &
Soft trilinear parameters $A^u_{ij}$ \\
\texttt{Ad} & \texttt{(3, 3)} &
Soft trilinear parameters $A^d_{ij}$ \\
\texttt{Ae} & \texttt{(3, 3)} &
Soft trilinear parameters $A^e_{ij}$ \\
\texttt{Av} & \texttt{(3, 3)} &
Soft trilinear parameters $A^\nu_{ij}$ \\
\texttt{Alam} & \texttt{(3, )} &
Soft trilinear parameters $A^\lambda_i$ \\
\texttt{Ak} & \texttt{(3, 3, 3)} &
Soft trilinear parameters $A^\kappa_{ijk}$ \\
\texttt{M1} &  &
Gaugino mass parameter $M_1$ \\
\texttt{M2} &  &
Gaugino mass parameter $M_2$ \\
\texttt{M3} &  &
Gaugino mass parameter $M_3$ \\
\texttt{CTW} &  &
Cosine of weak mixing angle $c_w$ \\
\texttt{STW} &  &
Sine of weak mixing angle $s_w$ \\
\texttt{g1} &  &
$\mathrm{U}(1)_Y$ gauge coupling $g_1$ \\
\texttt{g2} &  &
$\mathrm{SU}(2)_L$ gauge coupling $g_2$ \\
\texttt{g3} &  &
$\mathrm{SU}(3)_c$ gauge coupling $g_3$ \\
\texttt{v} &  &
SM vev $v$ \\
\texttt{vd} &  &
Down-type vev $v_d$ \\
\texttt{vu} &  &
Up-type vev $v_u$ \\
\texttt{Yu} & \texttt{(3, 3)} &
Up-type quark Yukawa couplings $Y^u_{ij}$ \\
\texttt{Yd} & \texttt{(3, 3)} &
Down-type quark Yukawa couplings $Y^d_{ij}$ \\
\texttt{Ye} & \texttt{(3, 3)} &
Charged lepton Yukawa couplings $Y^e_{ij}$ \\
\texttt{Tlam} & \texttt{(3, )} &
Soft trilinear couplings $T^\lambda_i$ \\
\texttt{Tk} & \texttt{(3, 3, 3)} &
Soft trilinear couplings $T^\kappa_{ijk}$ \\
\texttt{Tu} & \texttt{(3, 3)} &
Soft trilinear couplings $T^u_{ij}$ \\
\texttt{Td} & \texttt{(3, 3)} &
Soft trilinear couplings $T^d_{ij}$ \\
\texttt{Te} & \texttt{(3, 3)} &
Soft trilinear couplings $T^e_{ij}$ \\
\texttt{Tv} & \texttt{(3, 3)} &
Soft trilinear couplings $T^\nu_{ij}$ \\
\texttt{MuDimSq} &  &
Squared renormalization scale $\mu_R^2$ \\
\texttt{MUE} &  &
Effective $\mu$ parameter: $\mu = \lambda_i v_{iR} / \sqrt{2}$ \\
\texttt{mHd2} &  &
Down-type Higgs mass parameter $m_{H_d}^2$ \\
\texttt{mHu2} &  &
Up-type Higgs mass parameter $m_{H_u}^2$ \\
\texttt{mv2} & \texttt{(3, 3)} &
Right-handed sneutrino mass parameters $(m_{\widetilde{\nu}}^2)_{ij}$ \\
\texttt{ml2} & \texttt{(3, 3)} &
Left-handed sneutrino and slepton mass parameters $(m_{\widetilde{L}}^2)_{ij}$ \\[1em]
\caption{Class attributes set for an instance of the class
\texttt{BenchmarkPointFromFile} during initialization.
The second column shows the shape of the objects of
the type \texttt{arrayQP}. If no shape is shown the object
is of type \texttt{numberQP}.}
\label{bpinitatt}
\end{longtable}
}

{\renewcommand{\arraystretch}{1.2}
\footnotesize
\begin{longtable}{p{0.12\textwidth}p{0.12\textwidth}p{0.62\textwidth}}
{\normalsize \texttt{self.calc\_tree\_level\_spectrum()}} & \\
\hline
\texttt{MassSt} & \texttt{(2, )} &
Stop masses $m_{\widetilde{t}_i}$ \\
\texttt{ZT} & \texttt{(2, 2)} &
Stop mixing matrix $Z^{\widetilde{t}}_{ij}$ \\
\texttt{MassSc} & \texttt{(2, )} &
Scalar charme quark masses $m_{\widetilde{c}_i}$ \\
\texttt{ZC} & \texttt{(2, 2)} &
Scalar charme quark mixing matrix $Z^{\widetilde{c}}_{ij}$ \\
\texttt{MassSu} & \texttt{(2, )} &
Scalar up quark masses $m_{\widetilde{u}_i}$ \\
\texttt{ZU} & \texttt{(2, 2)} &
Scalar up quark mixing matrix $Z^{\widetilde{u}}_{ij}$ \\
\texttt{MassSb} & \texttt{(2, )} &
Sbottom masses $m_{\widetilde{b}_i}$ \\
\texttt{ZB} & \texttt{(2, 2)} &
Sbottom mixing matrix $Z^{\widetilde{b}}_{ij}$ \\
\texttt{MassSs} & \texttt{(2, )} &
Scalar strange quark masses $m_{\widetilde{s}_i}$ \\
\texttt{ZS} & \texttt{(2, 2)} &
Scalar strange quark mixing matrix $Z^{\widetilde{s}}_{ij}$ \\
\texttt{MassSd} & \texttt{(2, )} &
Scalar down quark masses $m_{\widetilde{d}_i}$ \\
\texttt{ZD} & \texttt{(2, 2)} &
Scalar down quark mixing matrix $Z^{\widetilde{d}}_{ij}$ \\
\texttt{Masshh} & \texttt{(8, )} &
Neutral CP-even scalar masses $m_{h_i}$ \\
\texttt{ZH} & \texttt{(8, 8)} &
Neutral CP-even scalar mixing matrix $Z^h_{ij}$ \\
\texttt{MassAh} & \texttt{(8, )} &
Neutral CP-odd scalar masses $m_{A_i}$ (including the Goldstone boson) \\
\texttt{ZA} & \texttt{(8, 8)} &
Neutral CP-odd scalar mixing matrix $Z^A_{ij}$ \\
\texttt{MassHpm} & \texttt{(8, )} &
Charged scalar masses $m_{H^\pm_i}$ (including the Goldstone boson) \\
\texttt{ZP} & \texttt{(8, 8)} &
Charged scalar mixing matrix $Z^{H^\pm}_{ij}$ \\
\texttt{MassCha} & \texttt{(5, )} &
Charged fermion masses $m_{\chi^\pm_i}$ \\
\texttt{ZEL} & \texttt{(5, 5)} &
Left-handed charged fermion mixing matrix $Z^{\chi^\pm_L}_{ij}$ \\
\texttt{ZER} & \texttt{(5, 5)} &
Right-handed charged fermion mixing matrix $Z^{\chi^\pm_R}_{ij}$ \\
\texttt{MassChi} & \texttt{(10, )} &
Neutral fermion masses $m_{\chi^0_i}$ \\
\texttt{UV\_Re} & \texttt{(10, 10)} &
Real part of the neutral fermion mixing matrix $Z^{\chi^0}_{ij}$ \\
\texttt{UV\_Im} & \texttt{(10, 10)} &
Imaginary part of the neutral fermion mixing matrix $Z^{\chi^0}_{ij}$ \\[1em]
\caption{Class attributes set for an instance of the class
\texttt{BenchmarkPointFromFile} by the method
\texttt{calc\_tree\_level\_spectrum}.
The second column shows the shape of the objects of
the type \texttt{arrayQP}.}
\label{bpspecatt}
\end{longtable}
}

{\renewcommand{\arraystretch}{1.2}
\footnotesize
\begin{longtable}{p{0.12\textwidth}p{0.14\textwidth}p{0.12\textwidth}|
p{0.12\textwidth}p{0.14\textwidth}p{0.12\textwidth}}
{\normalsize \texttt{self.calc\_tree\_level\_couplings()}} & \\
\hline
\texttt{hhhh} & \texttt{(8, 8, 8, 8)} &
$\Gamma_{h_i h_j h_k h_l}$ &
\texttt{hhh} & \texttt{(8, 8, 8)} &
$\Gamma_{h_i h_j h_k}$ \\
\texttt{AAh} & \texttt{(8, 8, 8)} &
$\Gamma_{A_i A_j h_k}$ &
\texttt{AAAA} & \texttt{(8, 8, 8, 8)} &
$\Gamma_{A_i A_j A_k A_l}$ \\
\texttt{AAhh} & \texttt{(8, 8, 8, 8)} &
$\Gamma_{A_i A_j h_k h_l}$ &
\texttt{AXX} & \texttt{(8, 8, 8)} &
$\Gamma_{A_i H^\pm_j H^\mp_k}$ \\
\texttt{hXX} & \texttt{(8, 8, 8)} &
$\Gamma_{h_i H^\pm_j H^\mp_k}$ &
\texttt{AAXX} & \texttt{(8, 8, 8, 8)} &
$\Gamma_{A_i A_j H^\pm_k H^\mp_l}$ \\
\texttt{AhXX} & \texttt{(8, 8, 8, 8)} &
$\Gamma_{A_i h_j H^\pm_k H^\mp_l}$ &
\texttt{XXXX} & \texttt{(8, 8, 8, 8)} &
$\Gamma_{H^\pm_i H^\mp_j H^\pm_k H^\mp_l}$ \\
\texttt{hhXX} & \texttt{(8, 8, 8, 8)} &
$\Gamma_{h_i h_j H^\pm_k H^\mp_l}$ & & \\
\texttt{ChaChaA1} & \texttt{(5, 5, 8)} &
$\Gamma^-_{\chi^\pm_i \chi^\mp_j A_k}$ &
\texttt{ChaChaA2} & \texttt{(5, 5, 8)} &
$\Gamma^+_{\chi^\pm_i \chi^\mp_j A_k}$ \\
\texttt{ChaChah1} & \texttt{(5, 5, 8)} &
$\Gamma^-_{\chi^\pm_i \chi^\mp_j h_k}$ &
\texttt{ChaChah2} & \texttt{(5, 5, 8)} &
$\Gamma^+_{\chi^\pm_i \chi^\mp_j h_k}$ \\
\texttt{ChaChiX1} & \texttt{(5, 10, 8)} &
$\Gamma^-_{\chi^\pm_i \chi^0_j H^\mp_k}$ &
\texttt{ChaChiX2} & \texttt{(5, 10, 8)} &
$\Gamma^+_{\chi^\pm_i \chi^0_j H^\mp_k}$ \\
\texttt{ChiChaX1} & \texttt{(10, 5, 8)} &
$\Gamma^-_{\chi^0_i \chi^\mp_j H^\pm_k}$ &
\texttt{ChiChaX2} & \texttt{(10, 5, 8)} &
$\Gamma^+_{\chi^0_i \chi^\mp_j H^\pm_k}$ \\
\texttt{ChiChiA1} & \texttt{(10, 10, 8)} &
$\Gamma^-_{\chi^0_i \chi^0_j A_k}$ &
\texttt{ChiChiA2} & \texttt{(10, 10, 8)} &
$\Gamma^+_{\chi^0_i \chi^0_j A_k}$ \\
\texttt{ChiChih1} & \texttt{(10, 10, 8)} &
$\Gamma^-_{\chi^0_i \chi^0_j h_k}$ &
\texttt{ChiChih2} & \texttt{(10, 10, 8)} &
$\Gamma^+_{\chi^0_i \chi^0_j h_k}$ \\
\texttt{ChaChay1} & \texttt{(5, 5)} &
$\Gamma^-_{\chi^\pm_i \chi^\mp_j \gamma}$ &
\texttt{ChaChay2} & \texttt{(5, 5)} &
$\Gamma^+_{\chi^\pm_i \chi^\mp_j \gamma}$ \\
\texttt{ChaChaZ1} & \texttt{(5, 5)} &
$\Gamma^-_{\chi^\pm_i \chi^\mp_j Z}$ &
\texttt{ChaChaZ2} & \texttt{(5, 5)} &
$\Gamma^+_{\chi^\pm_i \chi^\mp_j Z}$ \\
\texttt{ChaChiW1} & \texttt{(5, 10)} &
$\Gamma^-_{\chi^\pm_i \chi^0_j W^\mp}$ &
\texttt{ChaChiW2} & \texttt{(5, 10)} &
$\Gamma^+_{\chi^\pm_i \chi^0_j W^\mp}$ \\
\texttt{ChiChaW1} & \texttt{(10, 5)} &
$\Gamma^-_{\chi^0_i \chi^\mp_j W^\pm}$ &
\texttt{ChiChaW2} & \texttt{(10, 5)} &
$\Gamma^+_{\chi^0_i \chi^\pm_j W^\pm}$ \\
\texttt{ChiChiZ1} & \texttt{(10, 10)} &
$\Gamma^-_{\chi^0_i \chi^0_j Z}$ &
\texttt{ChiChiZ2} & \texttt{(10, 10)} &
$\Gamma^+_{\chi^0_i \chi^0_j Z}$ \\
\texttt{ASbSb} & \texttt{(8, 2, 2)} &
$2 \Gamma_{A_i \widetilde{b}_j \widetilde{\overline b}_k}$ &
\texttt{ASsSs} & \texttt{(8, 2, 2)} &
$2 \Gamma_{A_i \widetilde{s}_j \widetilde{\overline s}_k}$ \\
\texttt{ASdSd} & \texttt{(8, 2, 2)} &
$2 \Gamma_{A_i \widetilde{d}_j \widetilde{\overline d}_k}$ &
\texttt{AStSt} & \texttt{(8, 2, 2)} &
$2 \Gamma_{A_i \widetilde{t}_j \widetilde{\overline t}_k}$ \\
\texttt{AScSc} & \texttt{(8, 2, 2)} &
$2 \Gamma_{A_i \widetilde{c}_j \widetilde{\overline c}_k}$ &
\texttt{ASuSu} & \texttt{(8, 2, 2)} &
$2 \Gamma_{A_i \widetilde{u}_j \widetilde{\overline u}_k}$ \\
\texttt{hSbSb} & \texttt{(8, 2, 2)} &
$12 \Gamma_{h_i \widetilde{b}_j \widetilde{\overline b}_k}$ &
\texttt{hSsSs} & \texttt{(8, 2, 2)} &
$12 \Gamma_{h_i \widetilde{s}_j \widetilde{\overline s}_k}$ \\
\texttt{hSdSd} & \texttt{(8, 2, 2)} &
$12 \Gamma_{h_i \widetilde{d}_j \widetilde{\overline d}_k}$ &
\texttt{hStSt} & \texttt{(8, 2, 2)} &
$12 \Gamma_{h_i \widetilde{t}_j \widetilde{\overline t}_k}$ \\
\texttt{hScSc} & \texttt{(8, 2, 2)} &
$12 \Gamma_{h_i \widetilde{c}_j \widetilde{\overline c}_k}$ &
\texttt{hSuSu} & \texttt{(8, 2, 2)} &
$12 \Gamma_{h_i \widetilde{u}_j \widetilde{\overline u}_k}$ \\
\texttt{AASbSb} & \texttt{(8, 8, 2, 2)} &
$12 \Gamma_{A_i A_j \widetilde{b}_k \widetilde{\overline b}_l}$ &
\texttt{AASsSs} & \texttt{(8, 8, 2, 2)} &
$12 \Gamma_{A_i A_j \widetilde{s}_k \widetilde{\overline s}_l}$ \\
\texttt{AASdSd} & \texttt{(8, 8, 2, 2)} &
$12 \Gamma_{A_i A_j \widetilde{d}_k \widetilde{\overline d}_l}$ &
\texttt{AAStSt} & \texttt{(8, 8, 2, 2)} &
$12 \Gamma_{A_i A_j \widetilde{t}_k \widetilde{\overline t}_l}$ \\
\texttt{AAScSc} & \texttt{(8, 8, 2, 2)} &
$12 \Gamma_{A_i A_j \widetilde{c}_k \widetilde{\overline c}_l}$ &
\texttt{AASuSu} & \texttt{(8, 8, 2, 2)} &
$12 \Gamma_{A_i A_j \widetilde{u}_k \widetilde{\overline u}_l}$ \\
\texttt{hhSbSb} & \texttt{(8, 8, 2, 2)} &
$12 \Gamma_{h_i h_j \widetilde{b}_k \widetilde{\overline b}_l}$ &
\texttt{hhSsSs} & \texttt{(8, 8, 2, 2)} &
$12 \Gamma_{h_i h_j \widetilde{s}_k \widetilde{\overline s}_l}$ \\
\texttt{hhSdSd} & \texttt{(8, 8, 2, 2)} &
$12 \Gamma_{h_i h_j \widetilde{d}_k \widetilde{\overline d}_l}$ &
\texttt{hhStSt} & \texttt{(8, 8, 2, 2)} &
$12 \Gamma_{h_i h_j \widetilde{t}_k \widetilde{\overline t}_l}$ \\
\texttt{hhScSc} & \texttt{(8, 8, 2, 2)} &
$12 \Gamma_{h_i h_j \widetilde{c}_k \widetilde{\overline c}_l}$ &
\texttt{hhSuSu} & \texttt{(8, 8, 2, 2)} &
$12 \Gamma_{h_i h_j \widetilde{u}_k \widetilde{\overline u}_l}$ \\
\texttt{XXZ} & \texttt{(8, 8)} &
$\Gamma_{H^\pm_i H^\mp_j Z}$ &
\texttt{XXy} & \texttt{(8, 8)} &
$\Gamma_{H^\pm_i H^\mp_j \gamma}$ \\
\texttt{XXZZ} & \texttt{(8, 8)} &
$\Gamma_{H^\pm_i H^\mp_j Z Z}$ &
\texttt{XXyZ} & \texttt{(8, 8)} &
$\Gamma_{H^\pm_i H^\mp_j \gamma Z}$ \\
\texttt{XStSb} & \texttt{(8, 2, 2)} &
$4 \Gamma_{H^\pm_i \widetilde{\overline t}_j
\widetilde{b}_k}$ &
\texttt{XScSs} & \texttt{(8, 2, 2)} &
$4 \Gamma_{H^\pm_i \widetilde{\overline c}_j
\widetilde{s}_k}$ \\
\texttt{XSuSd} & \texttt{(8, 2, 2)} &
$4 \Gamma_{H^\pm_i \widetilde{\overline{u}}_j
\widetilde{d}_j}$ \\[1em]
\caption{Class attributes set for an instance of the class
\texttt{BenchmarkPointFromFile} by the method
\texttt{calc\_tree\_level\_couplings}. The name of
each attribute is given by \texttt{cpl\_[1]\_[2]}
with \texttt{[1]} being the string given in the first column
and \texttt{[2]} being \texttt{Re} for the real part
and \texttt{Im} for the imaginary part of the couplings.
The second column shows the shape of the objects of
the type \texttt{arrayQP}.
The couplings of fermions to scalars are decomposed as
$\Gamma = \Gamma^- \omega^- + \Gamma^+ \omega^+$ with
$\omega^\pm = (1 \pm \gamma_5)/2$. The couplings of fermions
to vector bosons are decomposed as
$\Gamma_\mu = \Gamma^- \gamma_\mu \omega^- + \Gamma^+
    \gamma_\mu \omega^+$.}
\label{bpspecatt}
\end{longtable}
}

{\renewcommand{\arraystretch}{1.2}
\footnotesize
\begin{longtable}{p{0.12\textwidth}p{0.12\textwidth}p{0.66\textwidth}}
{\normalsize \texttt{self.calc\_one\_loop\_counterterms()}} & \\
\hline
\texttt{dZhh} & \texttt{(8, 8)} &
Field renormalization counterterms $\delta Z_{h_i h_j}$ \\
\texttt{dZAA} & \texttt{(8, 8)} &
Field renormalization counterterms $\delta Z_{A_i A_j}$ \\
\texttt{dTphi\_Re} & \texttt{(8, )} &
Real part of the tadpole counterterms in the gauge basis
$\mathrm{Re}(\delta T_{\phi_i})$ \\
\texttt{dTphi\_Im} & \texttt{(8, )} &
Imaginary part of the tadpole counterterms in the gauge basis
$\mathrm{Im}(\delta T_{\phi_i})$ \\
\texttt{dMW2} &  &
$W$-boson mass counterterm $\delta M_W^2$ \\
\texttt{dMZ2} &  &
$Z$-boson mass counterterm $\delta M_Z^2$ \\
\texttt{dmlHd2} & \texttt{(3, )} &
Soft mass parameter counterterms $(\delta m_{H_d \widetilde{L}}^2 )_i$ \\
\texttt{dml2Sum12} &  &
Soft mass parameter counterterm $\delta (m_{\widetilde{L}}^2)_{12}$ \\
\texttt{dml2Sum13} &  &
Soft mass parameter counterterm $\delta (m_{\widetilde{L}}^2)_{13}$ \\
\texttt{dml2Sum23} &  &
Soft mass parameter counterterm $\delta (m_{\widetilde{L}}^2)_{23}$ \\
\texttt{dmv2Sum12} &  &
Soft mass parameter counterterm $\delta (m_{\widetilde{\nu}}^2)_{12}$ \\
\texttt{dmv2Sum13} &  &
Soft mass parameter counterterm $\delta (m_{\widetilde{\nu}}^2)_{13}$ \\
\texttt{dmv2Sum23} &  &
Soft mass parameter counterterm $\delta (m_{\widetilde{\nu}}^2)_{23}$ \\
\texttt{dvL2} & \texttt{(3, )} &
Vev counterterms $\delta v_{iL}^2$ \\
\texttt{dvR2} & \texttt{(3, )} &
Vev counterterms $\delta v_{iR}^2$ \\
\texttt{dv2} &  &
Vev counterterm $\delta v^2$ \\
\texttt{dTanBe} &  &
Parameter counterterm $\delta \tan\beta$ \\
\texttt{dlam} & \texttt{(3, )} &
Superpotential parameter counterterms $\delta \lambda_i$ \\
\texttt{dkap} & \texttt{(3, 3, 3)} &
Superpotential parameter counterterms $\delta \kappa_{ijk}$ \\
\texttt{dYv} & \texttt{(3, 3)} &
Superpotential parameter counterterms $\delta Y^\nu_{ij}$ \\
\texttt{dTlam} & \texttt{(3, )} &
Soft parameter counterterms $\delta T^\lambda_i$ \\
\texttt{dTk} & \texttt{(3, 3, 3)} &
Soft parameter counterterms $\delta T^\kappa_{ijk}$ \\
\texttt{dM1} &  &
Gaugino mass parameter counterterm $\delta M_1$ \\
\texttt{dM2} &  &
Gaugino mass parameter counterterm $\delta M_2$ \\
\texttt{dM2phiphi\_Re} & \texttt{(8, 8)} &
Real part of neutral CP-even scalar mass matrix counterterms in gauge
basis $\mathrm{Re}(\delta M^2_{\phi_i \phi_j})$ \\
\texttt{dM2phiphi\_Im} & \texttt{(8, 8)} &
Imaginary part of neutral CP-even scalar mass matrix counterterms in gauge
basis $\mathrm{Im}(\delta M^2_{\phi_i \phi_j})$ \\
\texttt{dM2sigsig\_Re} & \texttt{(8, 8)} &
Real part of neutral CP-odd scalar mass matrix counterterms in gauge
basis $\mathrm{Re}(\delta M^2_{\sigma_i \sigma_j})$ \\
\texttt{dM2sigsig\_Im} & \texttt{(8, 8)} &
Imaginary part of neutral CP-odd scalar mass matrix counterterms in gauge
basis $\mathrm{Im}(\delta M^2_{\sigma_i \sigma_j})$ \\
\texttt{dM2hh\_Re} & \texttt{(8, 8)} &
Real part of neutral CP-even scalar mass matrix counterterms in tree-level
mass eigenstate basis $\mathrm{Re}(\delta M^2_{h_i h_j})$ \\
\texttt{dM2hh\_Im} & \texttt{(8, 8)} &
Imaginary part of neutral CP-even scalar mass matrix counterterms in tree-level
mass eigenstate basis $\mathrm{Im}(\delta M^2_{h_i h_j})$ \\
\texttt{dM2AA\_Re} & \texttt{(8, 8)} &
Real part of neutral CP-odd scalar mass matrix counterterms in tree-level
mass eigenstate basis $\mathrm{Re}(\delta M^2_{A_i A_j})$ \\
\texttt{dM2AA\_Im} & \texttt{(8, 8)} &
Imaginary part of neutral CP-odd scalar mass matrix counterterms in tree-level
mass eigenstate basis $\mathrm{Im}(\delta M^2_{A_i A_j})$ \\[1em]
\caption{Class attributes set for an instance of the class
\texttt{BenchmarkPointFromFile} by the method
\texttt{calc\_one\_loop\_counterterms}.
The second column shows the shape of the objects of
the type \texttt{arrayQP}. If no shape is shown the object
is of the type \texttt{numberQP}. The counterterms are
calculated including the UV divergent piece proportional
to $\Delta = 1/\varepsilon^{\rm UV}$ (see \refta{bpinitatt})
The exact definitions of the counterterms can be found
in \citere{Biekotter:2019gtq}.}
\label{bpcountersatt}
\end{longtable}
}

{\renewcommand{\arraystretch}{1.2}
\footnotesize
\begin{longtable}{p{0.12\textwidth}p{0.12\textwidth}p{0.66\textwidth}}
{\normalsize \texttt{self.calc\_one\_loop\_self\_energies(even,odd,p2\_Re,p2\_Im)}} & \\
\hline
\texttt{hhSERen\_Re} & \texttt{(8, 8)} &
Real part of renormalized one-loop neutral CP-even
self energies $\mathrm{Re}(\hat\Sigma_{h_i h_j}(p^2))$
with $\mathrm{Re}(p^2)$ given as \texttt{p2\_Re}
and $\mathrm{Im}(p^2)$ given as \texttt{p2\_Im} \\
\texttt{hhSERen\_Im} & \texttt{(8, 8)} &
See above, but the imaginary part
$\mathrm{Im}(\hat\Sigma_{h_i h_j}(p^2))$ \\
\texttt{AASERen\_Re} & \texttt{(8, 8)} &
Real part of renormalized one-loop neutral CP-odd
self energies $\mathrm{Re}(\hat\Sigma_{A_i A_j}(p^2))$
with $\mathrm{Re}(p^2)$ given as \texttt{p2\_Re}
and $\mathrm{Im}(p^2)$ given as \texttt{p2\_Im} \\
\texttt{AASERen\_Im} & \texttt{(8, 8)} &
See above, but the imaginary part
$\mathrm{Im}(\hat\Sigma_{h_i h_j}(p^2))$ \\[1em]
\caption{The method \texttt{calc\_one\_loop\_self\_energies}
of the class \texttt{BenchmarkPointFromFile} returns the
values of the renormalized neutral scalar one-loop
self energies at the given momentum. The returned
object is a dictionary with the keys given in
the first column. The first two keys are present
if \texttt{even=1} is chosen and the latter two
keys are present if \texttt{odd=1} is chosen.
The values belonging to each key are objects
of type \texttt{arrayQP} with the shape
given in the second column.}
\label{bprenoseatt}
\end{longtable}
}

{\renewcommand{\arraystretch}{1.2}
\footnotesize
\begin{longtable}{p{0.12\textwidth}p{0.12\textwidth}p{0.66\textwidth}}
{\normalsize \texttt{self.calc\_two\_loop\_self\_energies(p2\_Re,p2\_Im)}} & \\
\hline
\texttt{hhSERen\_Re} & \texttt{(8, 8)} &
Real part of renormalized neutral CP-even
self energies including corrections
beyon one-loop level $\mathrm{Re}(\hat\Sigma_{h_i h_j}(p^2))$
with $\mathrm{Re}(p^2)$ given as \texttt{p2\_Re}
and $\mathrm{Im}(p^2)$ given as \texttt{p2\_Im} \\
\texttt{hhSERen\_Im} & \texttt{(8, 8)} &
See above, but the imaginary part
$\mathrm{Im}(\hat\Sigma_{h_i h_j}(p^2))$ \\[1em]
\caption{The method \texttt{calc\_two\_loop\_self\_energies}
of the class \texttt{BenchmarkPointFromFile} returns the
values of the renormalized neutral scalar 
self energies at the given momentum including
higher-order corrections beyond one-loop level.
The returned object is a dictionary with the keys given in
the first column.
The values belonging to each key are objects
of type \texttt{arrayQP} with the shape
given in the second column.}
\label{bprenose2latt}
\end{longtable}
}

{\renewcommand{\arraystretch}{1.2}
\footnotesize
\begin{longtable}{p{0.12\textwidth}p{0.12\textwidth}p{0.66\textwidth}}
{\normalsize \texttt{self.calc\_loop\_masses(even=2,odd=1,accu=1.e-5,momentum\_mode=1)}} & \\
\hline
\texttt{Masshh\_L} & \texttt{(8, )} &
Loop-corrected CP-even scalar masses at
one-loop level $m_{h_i}^{(1)}$ \\
\texttt{ZH\_L\_Re} & \texttt{(8, 8)} &
Real part of loop-corrected CP-even scalar mixing matrix
at one-loop level $\mathrm{Re}(Z^{h,(1)}_{ij})$ \\
\texttt{ZH\_L\_Im} & \texttt{(8, 8)} &
Imaginary part of loop-corrected CP-even scalar mixing matrix
at one-loop level $\mathrm{Im}(Z^{h,(1)}_{ij})$ \\
\texttt{Masshh\_2L} & \texttt{(8, )} &
Loop-corrected CP-even scalar masses including
one-loop and higher-order corrections $m_{h_i}^{(2')}$ \\
\texttt{ZH\_2L\_Re} & \texttt{(8, 8)} &
Real part of loop-corrected CP-even scalar mixing matrix
including one-loop and higher-order corrections $\mathrm{Re}(Z^{h,(2')}_{ij})$ \\
\texttt{ZH\_L\_Im} & \texttt{(8, 8)} &
Imaginary part of loop-corrected CP-even scalar mixing matrix
including one-loop and higher-order corrections  $\mathrm{Im}(Z^{h,(2')}_{ij})$ \\
\texttt{MassAh\_L} & \texttt{(8, )} &
Loop-corrected CP-odd scalar masses at
one-loop level $m_{A_i}^{(1)}$ (including the goldstone boson) \\
\texttt{ZA\_L\_Re} & \texttt{(8, 8)} &
Real part of loop-corrected CP-odd scalar mixing matrix
at one-loop level $\mathrm{Re}(Z^{A,(1)}_{ij})$ \\
\texttt{ZA\_L\_Im} & \texttt{(8, 8)} &
Imaginary part of loop-corrected CP-odd scalar mixing matrix
at one-loop level $\mathrm{Im}(Z^{A,(1)}_{ij})$ \\[1em]
\caption{The method \texttt{calc\_loop\_masses}
calculates the loop-corrected neutral scalar
spectrum. For \texttt{even=1} the attributes
\texttt{Masshh\_L}, \texttt{ZH\_L\_Re} and
\texttt{ZH\_L\_Im} are set.
For \texttt{even=2} the attributes
\texttt{Masshh\_2L\_Re}, \texttt{ZH\_2L\_Re}
and \texttt{ZH\_2L\_Im} are set.
For \texttt{odd=1} the attributes
\texttt{MassAh\_L}, \texttt{ZA\_L\_Re} and
\texttt{ZA\_L\_Im} are set.
The values of each attribute are objects
of type \texttt{arrayQP} with the shape
given in the second column.}
\label{bplmassatt}
\end{longtable}
}

{\renewcommand{\arraystretch}{1.2}
\footnotesize
\begin{longtable}{p{0.28\textwidth}p{0.12\textwidth}p{0.1\textwidth}}
{\normalsize \texttt{self.calc\_effective\_couplings()}} & \\
\hline
\texttt{ScalarCpls.chuu} & \texttt{(8, )} &
$c_{h_i u \bar u}$ \\
\texttt{ScalarCpls.chdd} & \texttt{(8, )} &
$c_{h_i d \bar d}$ \\
\texttt{ScalarCpls.chbb} & \texttt{(8, )} &
$c_{h_i b \bar b}$ \\
\texttt{ScalarCpls.chll} & \texttt{(8, )} &
$c_{h_i l \bar l}$ \\
\texttt{ScalarCpls.chtautau} & \texttt{(8, )} &
$c_{h_i \tau \bar \tau}$ \\
\texttt{ScalarCpls.chVV} & \texttt{(8, )} &
$c_{h_i V V}$ \\
\texttt{ScalarCpls.chgg} & \texttt{(8, )} &
$c_{h_i g g}$ \\
\texttt{ScalarCpls.chyy} & \texttt{(8, )} &
$c_{h_i \gamma \gamma}$ \\
\texttt{ScalarCpls.chAZ} & \texttt{(8, 8)} &
$c_{h_i A_j Z}$ \\
\texttt{Scalarcpls.chXW} & \texttt{(8, 8)} &
$c_{h_i H^\pm_j W^\mp}$ \\
\texttt{PseudoscalarCpls.cAuu} & \texttt{(8, )} &
$c_{A_i u \bar u}$ \\
\texttt{PseudoscalarCpls.cAdd} & \texttt{(8, )} &
$c_{A_i d \bar d}$ \\
\texttt{PseudoscalarCpls.cAbb} & \texttt{(8, )} &
$c_{A_i b \bar b}$ \\
\texttt{PseudoscalarCpls.cAll} & \texttt{(8, )} &
$c_{A_i l \bar l}$ \\
\texttt{PseudoscalarCpls.cAtautau} & \texttt{(8, )} &
$c_{A_i \tau \bar \tau}$ \\
\texttt{PseudoscalarCpls.cAVV} & \texttt{(8, )} &
$c_{A_i V V}$ \\
\texttt{PseudoscalarCpls.cAgg} & \texttt{(8, )} &
$c_{A_i g g}$ \\
\pagebreak
\texttt{PseudoscalarCpls.cAyy} & \texttt{(8, )} &
$c_{A_i \gamma \gamma}$ \\
\texttt{Pseudocalarcpls.cAXW} & \texttt{(8, 8)} &
$c_{A_i H^\pm_j W^\mp}$ \\[1em]
\caption{The method \texttt{calc\_effective\_couplings}
calculates the effective coupling coefficients, i.e.,
the couplings normalized to the SM prediction, for
the neutral scalars. The couplings between $h_i A_j Z$,
$h_i H^\pm_j W^\mp$ and $A_i H^\pm W^\mp$ do not have
an analogue in the SM. Instead, $c_{h_i A_j Z}$ is given
in factors of $e/(2 s_w c_w)$, and $c_{h_i H^\pm_j W^\mp}$
and $c_{A_i H^\pm_j W^\mp}$ in factors of $e/(2 s_w)$.
The objects \texttt{ScalarCpls} and \texttt{PseudoscalarCpls}
are set as attributes of the instance of \texttt{BenchmarkPointFromFile}.
They are themselves instances of the classes \texttt{Scalars}
and \texttt{Pseudoscalars} defined in the modules \texttt{scalars}
and \texttt{pseudoscalars} of the subpackage
\texttt{effectiveCouplings}. Thus, the first column shows the
attributes of the instance of \texttt{BenchmarkPointFromFile}.
They are \texttt{NumPy} arrays containing floats, with the
shape given in the second column.}
\label{bpeffcatt}
\end{longtable}
}

{\renewcommand{\arraystretch}{1.2}
\footnotesize
\begin{longtable}{p{0.24\textwidth}p{0.18\textwidth}p{0.18\textwidth}p{0.24\textwidth}}
{\normalsize \texttt{self.calc\_branching\_ratios()}} & \\
\hline
\texttt{Gammash[i]} & \texttt{hChiChi} & \texttt{(10, 10)} &
$\Gamma(h_i \rightarrow \chi^0_j \chi^0_k)$ \\
  & \texttt{hChaCha} & \texttt{(5, 5)} &
$\Gamma(h_i \rightarrow \chi^\pm_j \chi^\mp_k)$ \\
  & \texttt{hbb} &  &
$\Gamma(h_i \rightarrow b \bar b)$ \\
  & \texttt{htt} &  &
$\Gamma(h_i \rightarrow t \bar t)$ \\
  & \texttt{hcc} &  &
$\Gamma(h_i \rightarrow c \bar c)$ \\
  & \texttt{hss} &  &
$\Gamma(h_i \rightarrow s \bar s)$ \\
  & \texttt{hgg} &  &
$\Gamma(h_i \rightarrow g g)$ \\
  & \texttt{hyy} &  &
$\Gamma(h_i \rightarrow \gamma \gamma)$ \\
  & \texttt{hZZ} &  &
$\Gamma(h_i \rightarrow Z Z)$ \\
  & \texttt{hWW} &  &
$\Gamma(h_i \rightarrow W^\pm W^\mp)$ \\
  & \texttt{hhh} & \texttt{(8, 8)} &
$\Gamma(h_i \rightarrow h_j h_k)$ \\
  & \texttt{hAA} & \texttt{(7, 7)} &
$\Gamma(h_i \rightarrow A_j A_k)$ \\
  & \texttt{hAZ} & \texttt{(7, )} &
$\Gamma(h_i \rightarrow A_j Z)$ \\
  & \texttt{hXW} & \texttt{(7, 2)} &
$\Gamma(h_i \rightarrow H^{\pm,\mp}_j W^{\pm,\mp})$ \\
  & \texttt{hXX} & \texttt{(7, 7)} &
$\Gamma(h_i \rightarrow H^\pm_j H^\mp_k)$ \\
  & \texttt{hStSt} & \texttt{(2, 2)} &
$\Gamma(h_i \rightarrow \widetilde{t}_j \widetilde{\bar t}_k)$ \\
  & \texttt{hScSc} & \texttt{(2, 2)} &
$\Gamma(h_i \rightarrow \widetilde{c}_j \widetilde{\bar c}_k)$ \\
  & \texttt{hSuSu} & \texttt{(2, 2)} &
$\Gamma(h_i \rightarrow \widetilde{u}_j \widetilde{\bar u}_k)$ \\
  & \texttt{hSbSb} & \texttt{(2, 2)} &
$\Gamma(h_i \rightarrow \widetilde{b}_j \widetilde{\bar b}_k)$ \\
  & \texttt{hSsSs} & \texttt{(2, 2)} &
$\Gamma(h_i \rightarrow \widetilde{s}_j \widetilde{\bar s}_k)$ \\
  & \texttt{hSdSd} & \texttt{(2, 2)} &
$\Gamma(h_i \rightarrow \widetilde{d}_j \widetilde{\bar d}_k)$ \\
  & \texttt{Tot} &  & $\Gamma^{\rm Tot}_{h_i}$ \\
\texttt{BranchingRatiosh[i]} & \texttt{hChiChi} & \texttt{(10, 10)} &
$\mathrm{Br}(h_i \rightarrow \chi^0_j \chi^0_k)$ \\
  & \texttt{hChaCha} & \texttt{(5, 5)} &
$\mathrm{Br}(h_i \rightarrow \chi^\pm_j \chi^\mp_k)$ \\
  & \texttt{hbb} &  &
$\mathrm{Br}(h_i \rightarrow b \bar b)$ \\
  & \texttt{htt} &  &
$\mathrm{Br}(h_i \rightarrow t \bar t)$ \\
  & \texttt{hcc} &  &
$\mathrm{Br}(h_i \rightarrow c \bar c)$ \\
  & \texttt{hss} &  &
$\mathrm{Br}(h_i \rightarrow s \bar s)$ \\
  & \texttt{hgg} &  &
$\mathrm{Br}(h_i \rightarrow g g)$ \\
  & \texttt{hyy} &  &
$\mathrm{Br}(h_i \rightarrow \gamma \gamma)$ \\
  & \texttt{hZZ} &  &
$\mathrm{Br}(h_i \rightarrow Z Z)$ \\
  & \texttt{hWW} &  &
$\mathrm{Br}(h_i \rightarrow W^\pm W^\mp)$ \\
  & \texttt{hhh} & \texttt{(8, 8)} &
$\mathrm{Br}(h_i \rightarrow h_j h_k)$ \\
  & \texttt{hAA} & \texttt{(7, 7)} &
$\mathrm{Br}(h_i \rightarrow A_j A_k)$ \\
  & \texttt{hAZ} & \texttt{(7, )} &
$\mathrm{Br}(h_i \rightarrow A_j Z)$ \\
  & \texttt{hXW} & \texttt{(7, 2)} &
$\mathrm{Br}(h_i \rightarrow H^{\pm,\mp}_j W^{\pm,\mp})$ \\
  & \texttt{hXX} & \texttt{(7, 7)} &
$\mathrm{Br}(h_i \rightarrow H^\pm_j H^\mp_k)$ \\
  & \texttt{hStSt} & \texttt{(2, 2)} &
$\mathrm{Br}(h_i \rightarrow \widetilde{t}_j \widetilde{\bar t}_k)$ \\
  & \texttt{hScSc} & \texttt{(2, 2)} &
$\mathrm{Br}(h_i \rightarrow \widetilde{c}_j \widetilde{\bar c}_k)$ \\
  & \texttt{hSuSu} & \texttt{(2, 2)} &
$\mathrm{Br}(h_i \rightarrow \widetilde{u}_j \widetilde{\bar u}_k)$ \\
  & \texttt{hSbSb} & \texttt{(2, 2)} &
$\mathrm{Br}(h_i \rightarrow \widetilde{b}_j \widetilde{\bar b}_k)$ \\
  & \texttt{hSsSs} & \texttt{(2, 2)} &
$\mathrm{Br}(h_i \rightarrow \widetilde{s}_j \widetilde{\bar s}_k)$ \\
  & \texttt{hSdSd} & \texttt{(2, 2)} &
$\mathrm{Br}(h_i \rightarrow \widetilde{d}_j \widetilde{\bar d}_k)$ \\
\texttt{GammasA[i]} & \texttt{AChiChi} & \texttt{(10, 10)} &
$\Gamma(A_i \rightarrow \chi^0_j \chi^0_k)$ \\
  & \texttt{AChaCha} & \texttt{(5, 5)} &
$\Gamma(A_i \rightarrow \chi^\pm_j \chi^\mp_k)$ \\
  & \texttt{Abb} &  &
$\Gamma(A_i \rightarrow b \bar b)$ \\
  & \texttt{Att} &  &
$\Gamma(A_i \rightarrow t \bar t)$ \\
  & \texttt{Acc} &  &
$\Gamma(A_i \rightarrow c \bar c)$ \\
  & \texttt{Ass} &  &
$\Gamma(A_i \rightarrow s \bar s)$ \\
  & \texttt{Agg} &  &
$\Gamma(A_i \rightarrow g g)$ \\
  & \texttt{Ayy} &  &
$\Gamma(A_i \rightarrow \gamma \gamma)$ \\
  & \texttt{AAh} & \texttt{(7, 8)} &
$\Gamma(A_i \rightarrow A_j h_k)$ \\
  & \texttt{AhZ} & \texttt{(8, )} &
$\Gamma(A_i \rightarrow h_h Z)$ \\
  & \texttt{AXW} & \texttt{(7, 2)} &
$\Gamma(A_i \rightarrow H^{\pm,\mp}_j W^{\pm,\mp})$ \\
  & \texttt{AXX} & \texttt{(7, 7)} &
$\Gamma(A_i \rightarrow H^\pm_j H^\mp_k)$ \\
  & \texttt{AStSt} & \texttt{(2, 2)} &
$\Gamma(A_i \rightarrow \widetilde{t}_j \widetilde{\bar t}_k)$ \\
  & \texttt{AScSc} & \texttt{(2, 2)} &
$\Gamma(A_i \rightarrow \widetilde{c}_j \widetilde{\bar c}_k)$ \\
  & \texttt{ASuSu} & \texttt{(2, 2)} &
$\Gamma(A_i \rightarrow \widetilde{u}_j \widetilde{\bar u}_k)$ \\
  & \texttt{ASbSb} & \texttt{(2, 2)} &
$\Gamma(A_i \rightarrow \widetilde{b}_j \widetilde{\bar b}_k)$ \\
  & \texttt{ASsSs} & \texttt{(2, 2)} &
$\Gamma(A_i \rightarrow \widetilde{s}_j \widetilde{\bar s}_k)$ \\
  & \texttt{ASdSd} & \texttt{(2, 2)} &
$\Gamma(A_i \rightarrow \widetilde{d}_j \widetilde{\bar d}_k)$ \\
  & \texttt{Tot} &  & $\Gamma^{\rm Tot}_{A_i}$ \\
\texttt{BranchingRatiosA[i]} & \texttt{AChiChi} & \texttt{(10, 10)} &
$\mathrm{Br}(A_i \rightarrow \chi^0_j \chi^0_k)$ \\
  & \texttt{AChaCha} & \texttt{(5, 5)} &
$\mathrm{Br}(A_i \rightarrow \chi^\pm_j \chi^\mp_k)$ \\
  & \texttt{Abb} &  &
$\mathrm{Br}(A_i \rightarrow b \bar b)$ \\
  & \texttt{Att} &  &
$\mathrm{Br}(A_i \rightarrow t \bar t)$ \\
  & \texttt{Acc} &  &
$\mathrm{Br}(A_i \rightarrow c \bar c)$ \\
  & \texttt{Ass} &  &
$\mathrm{Br}(A_i \rightarrow s \bar s)$ \\
  & \texttt{Agg} &  &
$\mathrm{Br}(A_i \rightarrow g g)$ \\
  & \texttt{Ayy} &  &
$\mathrm{Br}(A_i \rightarrow \gamma \gamma)$ \\
  & \texttt{AAh} & \texttt{(7, 8)} &
$\mathrm{Br}(A_i \rightarrow A_j h_k)$ \\
  & \texttt{AhZ} & \texttt{(8, )} &
$\mathrm{Br}(A_i \rightarrow h_h Z)$ \\
  & \texttt{AXW} & \texttt{(7, 2)} &
$\mathrm{Br}(A_i \rightarrow H^{\pm,\mp}_j W^{\pm,\mp})$ \\
  & \texttt{AXX} & \texttt{(7, 7)} &
$\mathrm{Br}(A_i \rightarrow H^\pm_j H^\mp_k)$ \\
  & \texttt{AStSt} & \texttt{(2, 2)} &
$\mathrm{Br}(A_i \rightarrow \widetilde{t}_j \widetilde{\bar t}_k)$ \\
  & \texttt{AScSc} & \texttt{(2, 2)} &
$\mathrm{Br}(A_i \rightarrow \widetilde{c}_j \widetilde{\bar c}_k)$ \\
  & \texttt{ASuSu} & \texttt{(2, 2)} &
$\mathrm{Br}(A_i \rightarrow \widetilde{u}_j \widetilde{\bar u}_k)$ \\
  & \texttt{ASbSb} & \texttt{(2, 2)} &
$\mathrm{Br}(A_i \rightarrow \widetilde{b}_j \widetilde{\bar b}_k)$ \\
  & \texttt{ASsSs} & \texttt{(2, 2)} &
$\mathrm{Br}(A_i \rightarrow \widetilde{s}_j \widetilde{\bar s}_k)$ \\
  & \texttt{ASdSd} & \texttt{(2, 2)} &
$\mathrm{Br}(A_i \rightarrow \widetilde{d}_j \widetilde{\bar d}_k)$ \\
\texttt{GammasX[i]} & \texttt{XhX} & \texttt{(8, 7)} &
$\Gamma(H^\pm_i \rightarrow h_j H^\pm_k)$ \\
  & \texttt{XAX} & \texttt{(7, 7)} &
$\Gamma(H^\pm_i \rightarrow A_j H^\pm_k)$ \\
  & \texttt{XStSb} & \texttt{(2, 2)} &
$\Gamma(H^\pm_i \rightarrow \widetilde{t}_j \widetilde{\bar b}_k)$ \\
  & \texttt{XScSs} & \texttt{(2, 2)} &
$\Gamma(H^\pm_i \rightarrow \widetilde{c}_j \widetilde{\bar s}_k)$ \\
  & \texttt{XSuSd} & \texttt{(2, 2)} &
$\Gamma(H^\pm_i \rightarrow \widetilde{u}_j \widetilde{\bar d}_k)$ \\
  & \texttt{XhW} & \texttt{(8, )} &
$\Gamma(H^\pm_i \rightarrow h_j W^\pm)$ \\
  & \texttt{XAW} & \texttt{(7, )} &
$\Gamma(H^\pm_i \rightarrow A_j W^\pm)$ \\
  & \texttt{XChaChi} & \texttt{(5, 10)} &
$\Gamma(H^\pm_i \rightarrow \chi^\pm_j \chi^0_k)$ \\
  & \texttt{Xtb} &  &
$\Gamma(H^\pm_i \rightarrow t \bar b )$ \\
  & \texttt{Xcs} &  &
$\Gamma(H^\pm_i \rightarrow c \bar s )$ \\
  & \texttt{Tot} &  &
$\Gamma^{\rm Tot}_{H^\pm_i}$ \\[1em]
\caption{The method \texttt{calc\_branching\_ratios} calculates
the decay widths and the branching ratios of the neutral and
charged scalars. The results are set as attributes \texttt{Gammah},
\texttt{BranchingRatioh}, \texttt{GammaA}, \texttt{BranchingRatioA},
\texttt{GammaX} and \texttt{BranchingRatiosX}, which are lists of
dictionaries, to the instance of the class
\texttt{BenchmarkPointFromFile}.
Each dictionary contains the decay widths or branching
ratios of a particle $h_i$, $A_i$ or $H^\pm_i$, where $i=1,8$ for
the CP-even scalars and $i=1,7$ for the pseudoscalars and the
charged scalars. The different final states of the decays as given
in the fourth column correspond to the key values of the dictionaries
given in the second column. The values corresponding to each key
are floats if no shape is shown in the third column, and \texttt{NumPy}
arrays with the given shape otherwise.}
\label{bpbrsatt}
\end{longtable}
}

\end{document}